\documentclass[preprint,review,12pt]{elsarticle}

\usepackage{amssymb}
\usepackage{amsmath,bm}
\usepackage{colortbl,booktabs}
\usepackage{tabularx}
\usepackage{threeparttable}
\usepackage{multicol,multirow}
\usepackage{subfigure}
\usepackage [font=footnotesize,labelfont=bf]{caption}
\usepackage{rotating}
\usepackage{tikz}
\usepackage{hyperref}

\usepackage{color}

\journal{Elsevier}

\begin{document}

\begin{frontmatter}

	\title{A weakly compressible SPH method for RANS simulation of wall-bounded turbulent flows}

	\author[unit1]{Feng Wang}
	\ead{feng.wang.aer@tum.de}
	\author[unit2]{Zhongguo Sun}
	\ead{sun.zg@xjtu.edu.cn}
	\author[unit1]{Xiangyu Hu\texorpdfstring{\corref{mycorrespondingauthor}}{}}
	\cortext[mycorrespondingauthor]{Corresponding author.}
	\ead{xiangyu.hu@tum.de}

	\address[unit1]{School of Engineering and Design, Technical University of Munich\\
		85748 Garching, Germany}
	\address[unit2]{School of Energy and Power Engineering, Xi'an Jiaotong University\\
		710049 Xi'an, China}

	\begin{abstract}
		This paper presents a Weakly Compressible Smoothed Particle Hydrodynamics (WCSPH) method for solving the two-equation Reynolds-Averaged Navier-Stokes (RANS) model.
		The turbulent wall-bounded flow with or without mild flow separation, a crucial flow pattern in engineering applications, yet rarely explored in the SPH community, is simulated.
		The inconsistency between the Lagrangian characteristic and RANS model, mainly due to the intense particle shear and near-wall discontinuity, is firstly revealed and addressed by the mainstream and nearwall improvements, respectively.
		The mainstream improvements, including Adaptive Riemann-eddy Dissipation (ARD) and Limited Transport Velocity Formulation (LTVF), address dissipation incompatibility and turbulent kinetic energy over-prediction issues.
		The nearwall improvements, such as the particle-based wall model realization, weighted near-wall compensation scheme, and constant $y_p$ strategy, improve the accuracy and stability of the adopted wall model, where the wall dummy particles are still used for future coupling of solid dynamics.
		Besides, to perform rigorous convergence tests, an level-set-based boundary-offset technique is developed to ensure consistent $y^+$ across different resolutions.
		The benchmark wall-bounded turbulent cases, including straight, mildly- and strongly-curved, and Half Converging and Diverging (HCD) channels are calculated.
		Good convergence is, to our best knowledge, firstly achieved for both velocity and turbulent kinetic energy for the SPH-RANS method.
		All the results agree well with the data from the experiments or simulated by the Eulerian methods at engineering-acceptable resolutions.
		The proposed method bridges particle-based and mesh-based RANS models, providing adaptability for other turbulence models and potential for turbulent fluid-structure interaction (FSI) simulations.
	\end{abstract}


\end{frontmatter}
%
%
\section{Introduction}
\label{sec1}
The modern fluid machinery, including the turbulent chemical reactor\cite{manzano2017micromixing} and turbo-machinery\cite{wang2022simulation}, 
generally involve the problem of fluid-structure interaction (FSI).
The particle-based numerical methods, 
including the SPH (Smoothed Particle Hydrodynamics) 
and MPS (Moving Particle Semi-implicit) methods, 
have great potential at solving multi-physics, especially FSI, problems
\cite{zhang2021multi,liu2015particle,wang2022simulation,
	khayyer2018enhanced,morikawa2021coupling,liu2013improved}. 
Because the strong coupling can be easily achieved for all physical processes, 
and the fluid and solid dynamics equations can be discretized and solved within a unified computational framework \cite{zhang2020sphinxsys}.
However, without appropriate approach for simulating turbulent flows, 
the application of particle-based methods to practical engineering problems would be very limited,
since many of the industrial flows are highly turbulent.

The literature on turbulent flow simulations using particle-based methods 
generally falls into two categories. 
First, particle-based methods can inherently reproduce turbulence 
by solving the full Navier-Stokes equations, as seen in Direct Numerical Simulation (DNS) \cite{mayrhofer2015dns, bao2023pof}. 
But this is impractical for industrial applications
just as the traditional Computational Fluid Dynamics (CFD). 
Besides, the combination with Large Eddy Simulation (LES) is included in this category, as it primarily introduces the explicit 
\cite{dalrymple2006numerical, shao2012improved, ren2014numerical, duan2015large} 
or implicit \cite{monaghan2011turbulence, hu2015sph} sub-particle dissipations, without solving additional transport equations. 
The particle-based LES simulations effectively capture turbulent properties, 
such as kinetic energy near breaking waves \cite{dalrymple2006numerical}, 
porous structures \cite{ren2014numerical}, 
and phase interfaces \cite{duan2015large}. 
However, they struggle with wall-bounded turbulent flows 
and regions of strong shear stress, 
deviating significantly from reference data 
due to the lack of sufficient near wall resolution
\cite{arai2013large, shao2006sph}.

Second, the integration of RANS models, 
ranging from zero-equation \cite{nakayama2022wall, shao2012improved} 
to one-equation \cite{monaghan2011turbulence} 
and two-equation models \cite{wang2020isph, wang2022simulation}, 
with particle-based methods is 
theoretically much less demanding on near wall resolution.
With quite straightforward particle discretization of the RANS equations 
and implementation of the wall model similar to that of standard Eulerian mesh-based methods, 
most studies focus on ambient turbulent flows, 
such as breaking waves \cite{shao2005turbulence, shao2006simulation}, 
ocean currents \cite{shao2012improved, gotoh2004sph}, 
and near-shore dynamics \cite{lo2002simulation, dalrymple2006numerical}. 
While limited successes have been achieved, 
the particle-based RANS simulations generally over-predict 
the turbulent kinetic energy \cite{violeau2007numerical,wang2020isph}.
The production limiter, which is rarely used in the corresponding mesh-based cases, 
is adopted to mitigate the influence. 
However, the exact reason for the over-prediction remains unclear.
A further issue is 
that the wall-bounded turbulent flows remain largely unexplored, 
and even standard RANS benchmark cases have been merely reported, 
highlighting a critical gap in the research.

While the exact reason of such absence is undetermined,
we believe that an important difficulty origins from
the near wall modeling complicated by 
the Lagrangian characteristics of particle-based methods.
Nakayama et al. \cite{nakayama2022wall} 
developed a weakly compressible SPH (WCSPH) method 
for zero-equation RANS simulation of the turbulent open channel flow.
Other than the typical implementation of ghost- 
or dummy-particle methods to model the solid wall,
a "wall-layer" is introduced to move several layers of particles 
according to the wall model to eliminate 
the strong shear force resultant form the tangential velocity jump.
While being able to obtain correct near wall mean velocity profile, 
it cuts off the momentum exchange between the flow and solid wall, 
and hence limits the application to FSI problems \cite{zhang2021multi}.
Bao et al.\cite{bao2023pof} also reported a similar WCSPH method 
for $k-\epsilon$ RANS simulations of free stream and internal turbulent flows.
Different from Ref. \cite{nakayama2022wall}, 
a straightforward treatment like other previous works is 
applied to the near wall region with the compensation of very high spatial resolution. 
Therefore, even simple 2D RANS simulations are carried out with Graphic Process Unit (GPU) acceleration.  

In this paper, a WCSPH method is proposed for  $k-\epsilon$ RANS simulations of 
wall-bounded turbulent flow.
Apart from employing the state-of-the-art numerical schemes, 
including low-dissipative Riemann solver, Transport Velocity Formulation (TVF) and Reverse Kernel Gradient Correction (RKGC), 
to ensure stable and accurate discretization of the RANS equations,
the standard dummy-particles wall boundary, different from  Ref. \cite{nakayama2022wall}, 
is still used as aiming for future FSI application. 

Importantly, the present method addresses several critical challenges 
on the main-stream and near-wall treatments due to the Lagrangian characteristics 
of particle-based methods. 
First, an adaptive Riemann-eddy dissipation is proposed to 
ensure numerical stability in side stream
and to avoid over-damping in the middle stream.
Second, the kinetic energy over-prediction problem is linked with 
the TVF artificial production in the plug flow region,
and solved by a limited TVF technique.
Third, a Lagrangian-meshless implementation of 
the Eulerian-mesh originated wall model is proposed 
to handle the definition and adaptation of first fluid layer location $y_{p}$, 
the wall-function correction and the near-wall shear stress formulation.
Further more, to achieve SPH convergence of the $k-\epsilon$ RANS model, 
a constant $y_{p}$ strategy has been proposed together 
with corresponding boundary offset refinement.

As demonstrated in the numerical examples, 
the present method notably decrease 
the demanding of the near-wall spatial resolution,
and is able to achieve, for the first time to our knowledge,
rigorously validation of the SPH-RANS simulation with 
standard benchmark cases on wall-bounded turbulent flows.
The remainder of this manuscript is organized as follows.
Section \ref{section-wcsph-rans} introduces the governing equations 
and numerical discretization of the RANS equations.
The main-stream improvements and near wall treatment are respectively described in Section \ref{inner-region} and \ref{section-nearwall-treatment}.
Numerical examples are tested and discussed in Section \ref{section-numerical-examples}, 
and the concluding remarks are given in Section \ref{section-conclusion}.
The computational code of this work is released 
in the open-source SPHinXsys repository at https://github.com/Xiangyu-Hu/SPHinXsys.
%
%
\section{RANS turbulence model and WCSPH discretization}
\label{section-wcsph-rans}
\subsection{Governing equation}
Under the assumption of weakly compressible flow, 
the RANS model conservation equations of mass and momentum for 
the turbulent mean flow in the Lagrangian framework are
\begin{equation}
	\frac{\text{d} \rho}{\text{d} t} =  -\rho \nabla\cdot \mathbf v,
	\label{mass-equ}
\end{equation}
\begin{equation}
	\frac{\text{d} \mathbf v}{\text{d} t} =  - \frac{1}{\rho} \nabla p +  \mathbf g + \nu_l\nabla^2\mathbf v+\nabla\cdot\bm{\tau_t},
	\label{momentum-equ}
\end{equation}
where $\mathbf{v}$ is the flow velocity,
$\frac{\text d}{\text d t}=\frac{\partial}{\partial t} + \mathbf v \cdot \nabla$ 
stands for the material derivative,
$p$ is the pressure, $\rho$ is the density and $\nu_l$ is the molecular kinematic viscosity. 
To ensure small density variation, a stiff isothermal equation of state 
\begin{equation}
p = \rho_0 c^2_0 \left( \frac{\rho}{\rho_0} - 1\right),
\label{eos}
\end{equation}
where $\rho_0$ and $c_0$ are the reference density and 
a sufficient large speed of sound, respectively, is used.
$\bm{\tau_t}$ is the Reynolds stress tensor\cite{wilcox1998turbulence}, 
which can be calculated based on the Boussinesq approximation
\begin{equation}
	\bm{\tau_t} =2\nu_t \mathbf S - \frac{2}{3} k \mathbf I,
	\label{Boussinesq-equ}
\end{equation}
where $\nu_t$ is the kinematic eddy viscosity, $\mathbf S$ is the strain rate, $k$ is turbulent kinetic energy.
The strain rate can be expressed as $\mathbf S=(\nabla\mathbf v+\nabla\mathbf v^T)/2$. Substituting Eq. \eqref{Boussinesq-equ} into Eq. \eqref{momentum-equ} and considering the incompressible condition, the momentum equation can be rewritten as
\begin{equation}
	\frac{\text{d} \mathbf v}{\text{d} t} =  -  \frac{1}{\rho}\nabla p_{eff} +  \mathbf g +\nabla\cdot (\nu_{eff}\nabla \mathbf v),
	\label{momentum2-equ}
\end{equation}
where $ p_{eff}=p+(2/3) \rho k$ is the effective pressure 
and $ \nu_{eff} =\nu_{l} + \nu_{t} $ is the effective kinematic viscosity.
Note that, the $k$ contribution to $p_{eff}$ is usually 
taken into account implicitly in the incompressible projection method \cite{menter1994two}.
However, for the weakly compressible model used here, this contribute is applied explicitly.
Hereafter, the effective pressure and viscosity are denoted 
with the subscript $_{eff}$ omitted for simplicity.
 
To close the conservation equations, 
the two-equation $k - \epsilon$ model \cite{launder1983numerical} is adopted 
to determine $\nu_t$ and $k$ in Eq. (\ref{Boussinesq-equ}) by 
\begin{equation}
\nu_t= C_{\mu} \frac{k^2}{\epsilon},
\label{eddy-viscosity-equ}
\end{equation}
and the transport equations of $k$ and the turbulence dissipation rate $\epsilon$ as
\begin{equation}
	\frac{\text{d} k}{\text{d} t} = G_k -\epsilon+\nabla\cdot (D_k\nabla k),
	\label{k-equ}
\end{equation}
\begin{equation}
	\frac{\text{d} \epsilon}{\text{d} t} = C_1 \frac{\epsilon}{k} G_k -C_2 \frac{\epsilon^2}{k}+\nabla\cdot (D_\epsilon\nabla \epsilon),
	\label{epsilon-equ}
\end{equation}
where $D_k = \nu_l+\nu_t/\sigma_k$ and $D_\epsilon = \nu_l+\nu_t/\sigma_\epsilon$ 
are the diffusion coefficients for $k$ and $\epsilon$, respectively. 
Here, $C_{\mu}$, $C_1$, $C_2$, $\sigma_k$ and $\sigma_\epsilon$ are the empirical constants.
Note that, the 3 terms on the right hand sides (RHS), from left to right, of the two equations can be regarded as the production, dissipation and diffusion terms, respectively,
and the production term of the turbulent kinetic energy is defined as 
\begin{equation}
	G_k=\bm{\tau_t}\nabla \mathbf{v}.
	\label{production-equ}
\end{equation}
\subsection{Wall model}
\label{wall-model}
The wall model approximates the near-wall turbulence according to the law of wall 
obtained from numerous experiments. 
Without resolving the near-wall flow profile explicitly, 
the wall shear stress, 
turbulent kinetic energy production and turbulent dissipation rate 
at the location adjacent to the wall are determined approximately
according to the wall function.
In this work, the step-wise (standard) wall function 
\cite{launder1983numerical, spalart2000strategies}
is adopted because of its broad applicability, 
good convergence and high computational efficiency.

With the known or prescribed distance $y_p$ to the wall,
where the subscript $_p$ refers to a quantity at the wall-adjacent location,
the dimensionless wall distance $y^+$ is introduced as
\begin{equation}
y^+=\frac{y_p C^{1/4}_\mu k^{1/2}_p}{\nu_l}.
\label{yplus-equ2}
\end{equation}
The step-wise (standard) wall function,
relating the dimensionless velocity $u^+$ and $y^+$, hence is given as
\begin{equation}
u^+=
\begin{cases} 
\dfrac{1}{\kappa} \ln\left(E y^+\right), & y^+ > 11.225, \\[10pt]
y^+, & y^+ < 11.225,
\end{cases}
\label{eq:velo_grad_wall}
\end{equation}
where $E=0.9$ and $\kappa=0.41$ are the empirical constants.
Note that, Eq. (\ref{eq:velo_grad_wall}) has 
a limited ability to address the small 
but non-turbulent flow region
as $y^+ =11.225 $ serves as a criterion to identify 
the fully-turbulent (logarithmic) region and the laminar (linear) one. 
Also note that, 
the effective range of $y^+ \sim 30-100$, 
in which the wall function is able to address full developed turbulent flow \cite{spalart2000strategies},
suggests that $y_p$ is a not free parameter and should be chosen carefully so that 
$y^+$ falls into the effective range.

According to the kinetic-based formulation \cite{spalart2000strategies}, 
the wall shear stress is determined as 
\begin{equation}
    \tau_w =\frac{\rho U_p C^{1/4}_\mu k^{1/2}_p}{u^+} 
    \label{eq:wall-shear-stress}
\end{equation}
where $U_p$ is the wall-adjacent tangential velocity 
obtained from the moment equation Eq. (\ref{momentum-equ}) or (\ref{momentum2-equ}).
Other than obtained from Eq. (\ref{production-equ}), 
the production of turbulent kinetic energy at $y_p$ is obtained as
\begin{equation}
	G_{k,p}=\tau_{w} (\frac{\partial U}{\partial n})_{p},
	\label{wallfunc-production-k-equ2}
\end{equation}
where $(\frac{\partial U}{\partial n})_{p}$ refers to the velocity gradient normal to the wall surface, 
and can be computed by taking the derivative of the wall function, i.e. Eq. \eqref{eq:velo_grad_wall}.
Similarly, other than obtained from the transport equation Eq. (\ref{epsilon-equ}),
the dissipation rate at $y_p$ is obtained as
\begin{equation}
	\epsilon_p=\frac{C^{3/4}_{\mu}k^{3/2}_P}{\kappa y_p},
	\label{epsilon-at-P-equ2}
\end{equation}
according to the assumption of local equilibrium 
in the turbulence boundary layer \cite{launder1983numerical}.
Note that, such assumption is in agreement with 
the previous mentioned effective range of $y^+$ and 
the discussion of proper choice of $y_p$.
\subsection{WCSPH discretization}
\subsubsection{Discretization of conservation equations}
The conservation equations are discretized by the WCSPH formulation 
based on a low-dissipative Riemann solver \cite{zhang2017weakly} to increase stability.
The discretization of continuity equation at particle $i$ is
\begin{equation}
	\frac{\text{d} \rho_i}{\text{d} t} = 2 \rho_i \sum_{j} (\mathbf{v}_i-\mathbf{v^*}) 
	\cdot \nabla W_{ij} V_j,
\label{discretize-continuity-equ-riemann}
\end{equation}
where $V_j$ is the volume of neighbor particles, 
$\mathbf{v^*}= U^* \mathbf{e}_{ij}+(\overline{\mathbf{v}}_{ij} - \overline{U}_{ij}\mathbf{e}_{ij})$
is the intermediate velocity. 
This expression ensures that dissipation is applied exclusively in the pairwise direction
denoted as the unit vector  $\mathbf{e}_{ij}$.
Here, the gradient of the kernel function is expressed as 
$\nabla W_{ij} = \frac{\partial W_{ij}}{\partial r_{ij}} \mathbf{e}_{ij}$, 
where $W_{ij}$ represents $W(\mathbf{r}_{ij}, h)$ 
and $h$ is the smoothing length that is fixed at 1.3$dp$, where  $dp$ is particle spacing.
$\overline{(\bullet)}_{ij} = [(\bullet)_i + (\bullet)_j]/2$ means the pairwise average and $(\bullet)_{ij} = (\bullet)_i - (\bullet)_j$ the pairwise difference.
$\overline{U}_{ij} = \overline{\mathbf{v}}_{ij} \cdot \mathbf{e}_{ij}$ 
is the projection of pairwise-average velocity along the pairwise direction,
and the intermediate velocity is calculated by
\begin{equation}
	U^*=\overline{U}_{ij} +\frac{p_{ij}}{2\rho_0 c_0}.
	\label
	{intermediate-vel}
\end{equation}
For the momentum equation, 
the acceleration contributed by the pressure gradient is
\begin{equation}
	\left(\frac{\text{d} \mathbf v_i}{\text{d} t}\right)^p = -2\sum_{j} m_j \frac{p^*}{\rho_i\rho_j} \nabla W_{ij}  ,
	\label{discretize-momentum-p-intermediate}
\end{equation}
where $p^*$ is the intermediate pressure is given as
\begin{equation}
p^* = \overline{p}_{ij} +  \frac{1}{2}\beta_{ij}\rho_0 c_0 U_{ij}.
\label{discretize-momentum-p-intermediate-component}
\end{equation}
where $\beta_{ij} = \min( \eta \max(\mathbf{v}_{ij} \cdot\mathbf{e}_{ij},0), c_0 )$ 
is the dissipation limiter, and $\eta = 3$ is a generally effective empirical parameter.
Note that, the intermediate pressure contains two parts, 
the pairwise-average one and the dissipative one 
obtained from the Riemann solver\cite{zhang2017weakly}.
By employing the RKGC scheme \cite{zhang2024towards} to the pairwise-average term 
to increase the consistency and accuracy of this conservative operator,
the final form of Eq. \eqref{discretize-momentum-p-intermediate} can be expressed as  
\begin{equation}
	\left(\frac{\text{d} \mathbf v_i}{\text{d} t}\right)^p = -\sum_{j} m_j 
	\left(
	\frac{p_i \mathbf{B}_j+p_j \mathbf{B}_i}{\rho_i\rho_j} +
	\beta_{ij}\frac{ \rho_0 c_0 U_{ij}}{\rho_i\rho_j}\mathbf{I}
	\right)	\nabla W_{ij},
	\label{discretize-momentum-p-equ-riemann-final}
\end{equation}
where $\mathbf{I}$ is identity matrix and 
the correction matrix is calculated by
\begin{equation}
	\mathbf{B}_i = \left( - \sum_j \mathbf{r}_{ij} \otimes \nabla W_{ij} V_j \right)^{-1}.
	\label{B-matrix-equ}
\end{equation}

As for the viscous term in Eq. \eqref{momentum2-equ}, the strategy of firstly calculating the velocity gradient \cite{wang2022isph, lo2002simulation, gotoh2004sph, shao2006simulation}, then multiplying it by the effective viscosity, and finally computing its divergence, can result in the nesting issue \cite{hu2006multi,bao2023pof}.
To avoid the accuracy degradation induced by this problem, the compact formulation of the second-order derivative\cite{hu2006multi} is used.
The acceleration contributed by the viscous force is calculated as
\begin{equation}
	\left(\frac{\text{d} \mathbf v_i}{\text{d} t}\right)^\nu=2\sum_{j} m_j \widetilde{\mu}_{ij} \frac{\mathbf{v}_{ij}}{r_{ij}}\frac{\partial W_{ij}}{\partial r_{ij}},
	\label{turbu-viscous-equ}
\end{equation}
where $\widetilde{(\bullet)}_{ij} = 2(\bullet)_i  (\bullet)_j/[(\bullet)_i + (\bullet)_j]$ means the pairwise harmonic average for variable effective viscosity.
\subsubsection{Discretization of transport equations}
The velocity gradient, for the production term $G_k$ of Eq. (\ref{production-equ}) in 
both the $k$ and $\epsilon$ equations, 
is approximated with kernel correction as
\begin{equation}
	\nabla \mathbf{v}_i=\sum_{j} \mathbf{v}_{ij} \otimes (\mathbf{B}_i \nabla W_{ij}) V_j.
	\label{velo-grad-equ}
\end{equation}

Note that, since $k$ and $\epsilon$ equations are inherently non-conservative, the strong form gradient operator\cite{zhang2021sphinxsys} together with the kernel gradient correction is used to achieve good accuracy.
With the treatment of the diffusion terms similar to that for the viscous term, 
the discretization of $k$ and $\epsilon$ equations are given as 
\begin{equation}
	\frac{\text{d}  k_i}{\text{d} t}= \left(S_k\right)_i
	+2\sum_{j} m_j  \widetilde{(D_{k})}_{ij} 
	\frac{k_{ij}}{r_{ij}}\frac{\partial W_{ij}}{\partial r_{ij}},
	\label{k-discretized-equ}
\end{equation}
where $S_k =G_k -\epsilon $, and 
\begin{equation}
	\frac{\text{d} \epsilon_i}{\text{d} t} = 
	\left(S_\epsilon \right)_i
	+2\sum_{j} m_j \widetilde{(D_{\epsilon})}_{ij}  \frac{\epsilon_{ij}}{r_{ij}}\frac{\partial W_{ij}}{\partial r_{ij}},
	\label{epsilon-discretized-equ}
\end{equation}
where $S_\epsilon =  C_1 \epsilon G_k /k - C_2 \epsilon^2/ k$.

It is worthy noting that these above discretized equations 
only consider the contribution from the fluid or inner particles,
and the wall boundary conditions will be specifically 
considered in the section \ref{wall-function-sec}.
\subsection{Transport Velocity Formulation(TVF)}
For strong shear flow, 
the operator accuracy deteriorates\cite{dehnen2012improving} 
due to the degenerated particle distribution.
Here, the TVF technique \cite{adami2013transport, zhang2017generalized} 
is used to maintain the uniform particle distribution and improve the accuracy.
The latest version of this correction \cite{zhu2021consistency} 
is directly correlated with the zero-order consistency error.
In short, the position of the fluid particles is modified by
\begin{equation}
	\Delta \mathbf{r}_i = \alpha  h^2 \mathbf{R}_{\nabla \phi}^0.
	\label{tvf-orginal}
\end{equation}

Here, $\mathbf{R}_{\nabla \phi}^0$ is the zeroth-order consistency residue and calculated by
\begin{equation}
	\mathbf{R}_{\nabla \phi}^0=  \sum_j (\mathbf{B}_i + \mathbf{B}_j) \nabla W_{ij} V_j,
	\label{tvf-0-residue}
\end{equation}
where $\alpha = 0.2$ is the generally effective parameter chosen according to the time-step criteria\cite{zhang2024towards}.
\subsection{Duel-criteria time stepping and source term linearization}
The dual-criteria time stepping scheme\cite{zhang2020dual} is adopted to reduce computational costs, 
and the computation of $k$ and $\epsilon$ equations is included in the advection time step.
Compared with the original version, one additional modification is that the advection time criterion should be changed by including the effective viscosity as
\begin{equation}
\Delta t =\text{CFL}_{ad}\text{min}(h/\left\|\mathbf{v}\right\|_{max},h^2/\nu_{eff}).
\label{turbu-advec-time_equ}
\end{equation}

In the $k$ and $\epsilon$ transport equations, the source terms, including dissipation and production terms, are highly non-linear.
Therefore, the source term linearization scheme\cite{tao2001numerical} is adopted for stability.
The source terms in Eqs. (\ref{k-discretized-equ}) and (\ref{epsilon-discretized-equ}) 
are reconstructed according to the form of the linear function, $S_c+S_p\phi$, as
\begin{equation}
S_k = G_k^{n-1}-(\frac{\epsilon^{n-1}}{k^{n-1}} )k^n ,
\label{sl-k}
\end{equation}
\begin{equation}
S_{\epsilon} = C_1 \frac{\epsilon^{n-1}}{k^{n-1}} G_k^{n-1}-C_2 (\frac{\epsilon^{n-1}}{k^{n-1}} )\epsilon^n ,
\label{sl-ep}
\end{equation}
where $S_c$ and $S_p$ are the constant components of the source term
obtained from the previous time step $n-1$;
$\phi$ is the unknown variable such as $k^n$ and $\epsilon^n$ at the current time step $n$.
Introducing the current time step values into the dissipation term 
can improve the algorithm stability.
Because the minus operation will be optimized 
by moving this linearized term to the left hand side (LHS).
Taking the $k$ equation as the example, 
the explicit time stepping after linearization becomes
\begin{equation}
k^n = \frac{G_k^{n-1}\Delta t+ D_{k}\Delta t+k^{n-1}  }{1+\epsilon^{n-1}/k^{n-1}\Delta t}.
\label{sl-kn}
\end{equation}
Compared with the original formulation, Eq. \eqref{k-equ}, 
the linearized version consist of all positive operation 
and hence avoid the instability caused by the source term.
\section{Improving the simulation of main stream}
\label{inner-region}
\subsection{Adaptive Riemann-eddy dissipation}
\label{ARD}
\begin{figure}[htb!]
	\centering
	\includegraphics[trim = 0cm 0cm 10cm 11cm, clip,width=1.0\textwidth]{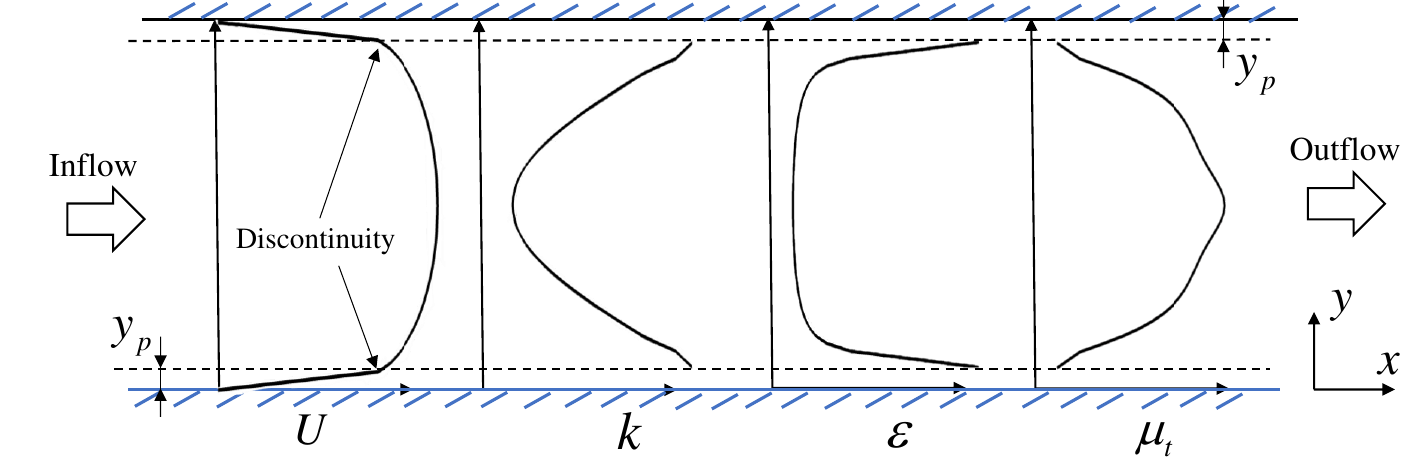}
	\caption{The profiles (from left to right) 
	of the mean flow velocity $U$, turbulent kinetic energy $k$, turbulent dissipation rate $\epsilon$ 
	and eddy viscosity $\mu_t$, obtained from $k-\epsilon$ RANS model in a fully-developed turbulent straight channel. Note the main stream and wall-adjacent region is separated by the dash lines indicating the first fluid layer thickness $y_p$.}
	\label{schemetic-rans}
\end{figure}

\begin{figure}[htb!]
	\centering
	\includegraphics[trim = 0cm 0cm 9cm 11cm, clip,width=0.9\textwidth]{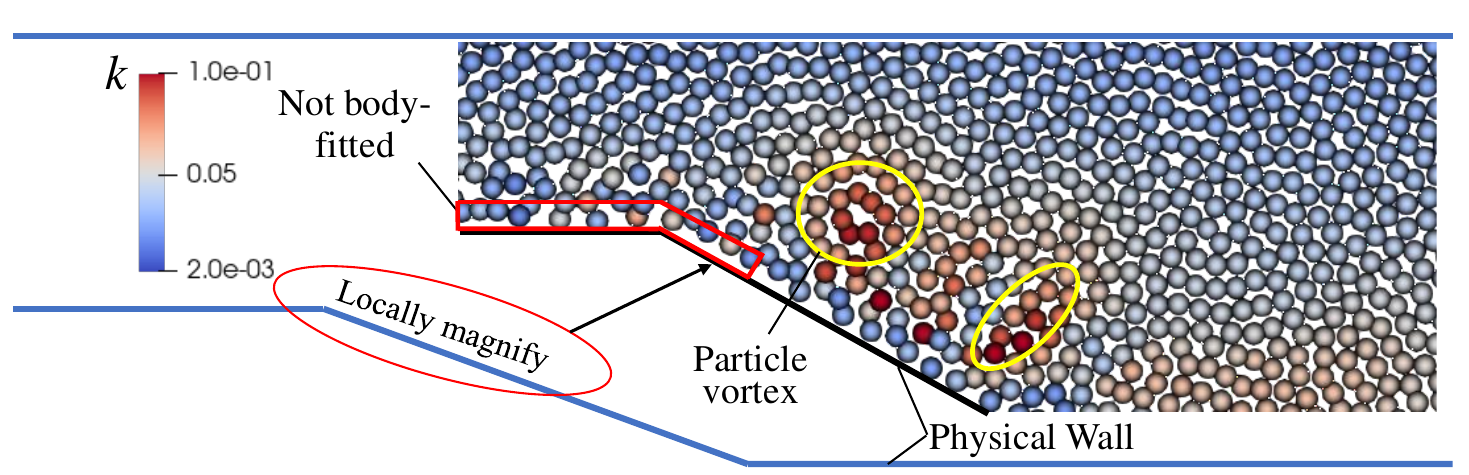}
	\caption{The typical flow field  in a turbulent expanding channel 
		obtained without Riemann dissipation.
	}
	\label{Problem_particle_cluster}
\end{figure}

As shown in the previous section, 
the Riemann-solver based numerical dissipation is sufficient 
to stabilize the high shear flows even for inviscid flows.
However, in the context of RANS turbulent flow, 
taking that in a straight channel case as an example, 
as shown in Fig. \ref{schemetic-rans}.
The Riemann dissipation may lead to over damping since it is superimposed on 
the dissipation effect of the eddy viscosity,
especially at the middle stream,
where the latter is already fairly high and the flow shear is mild.
In contrast, 
not applying the Riemann dissipation may lead to instability in the side stream
where higher shear rate is exhibited with fairly low eddy viscosity.
Another example is the flow region with adverse pressure gradient, 
as shown in Fig. \ref{Problem_particle_cluster}.
Without sufficient numerical dissipation, 
the particles are prone to cluster and formulate 
unsteady vortex-like pattern,
which is inconsistent with RANS model 
characterized with smooth velocity distribution 
except at wall-adjacent location.

These numerical vortexes may introduce extra production of the turbulent kinetic energy, and break the fundamental eddy-viscosity assumption, eventually reducing the accuracy of the mean flow properties.
Furthermore, the wall-nearest particles become not body-fitted, as shown in Figure \ref{Problem_particle_cluster} .
The strong shear effect makes them continuously migrate in and out of the near wall region.
This migration phenomenon not only deteriorates the stability, but also reduces the accuracy of the wall function.
Because some particles can be squeezed against the wall, causing an undesirable decrease in $y^+$ into the buffer-layer region where error accumulates.

In comparison, the mesh-based method such as the finite volume method does not have this problem.
Since it naturally has large dissipation near the boundary due to the cell flux average\cite{monaghan1992smoothed}.
And the wall-nearest grid is fixed to be body-fitted.

To handle such issue, one straightforward way is to switch between these
two dissipation mechanisms.
One can first approximate the force due 
to Riemann-dissipation 
in Eq. (\ref{discretize-momentum-p-intermediate-component}) as
\begin{equation}
(\frac{\text{d} \mathbf v}{\text{d} t})^{\mu_{R}}=\frac{2}{\rho_i}\sum_{j} \mu_{R}  \frac{\mathbf{v}_{ij}}{r_{ij}}\frac{\partial W_{ij}}{\partial r_{ij}}V_j,
\label{numer-viscous-equ}
\end{equation}
where $\mu_{R} = \frac{1}{2}\beta_{ij}\rho_0 c_0 h$ is the numerical viscosity,
by assuming the equivalence between smoothing length and particle distances.  
With a modulation function 
\begin{equation}
\mu_{c} =\max(\mu_{R}, \widetilde{\mu}_{ij}),
\label{extra-dissipation}
\end{equation}
the dissipative force switches between Eq. (\ref{turbu-viscous-equ}) 
and (\ref{numer-viscous-equ}) adaptively, 
and is able to provide sufficient dissipation without over-damping. 
\subsection{Limited TVF}
As shown in the last section, 
TVF is employed to increase the regularity of 
particle distribution 
and decrease corresponding numerical error.
Since the consistency residue is not able to be exactly zero 
due to the explicit steps of Eq. (\ref{tvf-orginal}), 
the correction continues and slightly modifies the particle positions
even in the flow region where the particle regularity is already quite good.
Generally, such tiny correction leads to no notable issue
due to the small residue given by Eq. (\ref{tvf-0-residue}).
However, it can result serious issue for the wall-bounded turbulent flow, 
especially for the channel or pipe flow, 
where a distinct plug flow pattern presents. 

In the plug flow region, 
for example the turbulent flow in a straight channel
as shown in Fig. \ref{schemetic-rans}, 
the exact solution gives a flow field with small, steady 
and unidirectional transverse velocity gradient.
The induced kinetic energy generated, as the fist term on right-hand-side of Eq. (\ref{k-equ}), 
is small and balanced with the turbulent dissipation.
However, because of the frequent TVF correction of particle position without modifying the velocity, 
considerable disturbance error can be introduced in the calculation of the velocity gradient.
According to the eddy-viscosity model, such error is quadratically correlated with the turbulent kinetic energy, as indicated in Eq. \eqref{production-equ}, 
suggesting that the tiny velocity gradient disturbance will be recognized 
as the "vortex" and hence produce extra kinetic energy, 
leading to the over-prediction problem.

Note that this issue becomes more pronounced as the resolution increases, 
because the higher resolution means the more correction iterations,
but the consistency residue of the WCSPH method can be saturated
(magnitude does not decrease beyond a threshold even with the higher resolution). 
And hence, for the original TVF, increasing resolution merely leads to the more accumulated disturbance error.

To handle this problem, we introduce a limited TVF with 
a linear limiter\cite{wang2024efficient} 
in the original formulation to restrict the over-correction as
\begin{equation}
	\beta_{tvf} = \min(mh^2 \left\| \mathbf{R}_{\nabla \phi}^0 \right\|^2, 1),
	\label{tvf-limiter}
\end{equation}
where $m=1000$ is the decaying slope employed in this work. 
The original correction Eq. \eqref{tvf-orginal} hence is modified as 
\begin{equation}
\Delta \mathbf{r}_i = \alpha  h^2 \beta_{tvf} \mathbf{R}_{\nabla \phi}^0.
\label{tvf-limited}
\end{equation}

\begin{figure}[htb!]
	\centering
	\includegraphics[trim = 0cm 0cm 8cm 2cm, clip,width=0.90\textwidth]{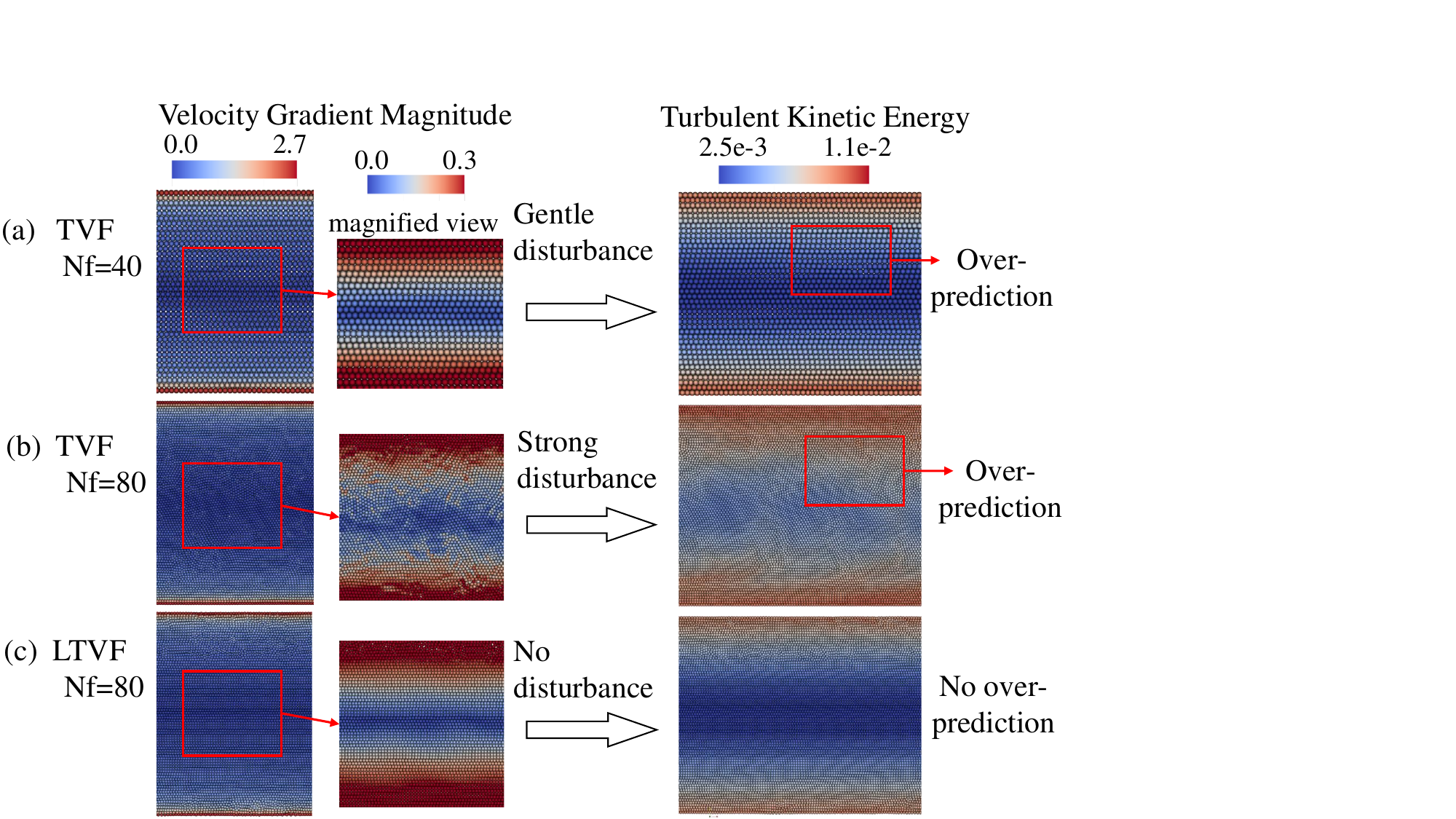}
	\caption{In a fully-developed turbulent straight channel, the results with the original 
	and limited TVF on velocity gradient and turbulent kinetic energy. $N_f$ gives the number of fluid particles on the cross-section.
	}
	\label{LTVF-effect-with-resolutions}
\end{figure}

Taking the fully-developed straight channel as the example, 
Fig. \ref{LTVF-effect-with-resolutions} (a) and \ref{LTVF-effect-with-resolutions} (b) 
show the results with the original TVF at the two resolutions.
The disturbance appears near the centerline (or the plug flow region) of the channel, 
and becomes more serious with increasing resolution.
Figure \ref{LTVF-effect-with-resolutions} (c) 
presents the result with the limited TVF, 
the disturbance on the velocity gradient, 
even at the high resolution, disappears 
and the over-prediction problem is well addressed.
%
%
\section{Near wall treatment and boundary condition}
\label{section-nearwall-treatment}
\subsection{Near-wall Lagrangian characteristics}
In RANS model, as shown in Fig. \ref{schemetic-rans},
the fluid domain is divided into the main stream
and a wall-adjacent unresolved region.
In the latter, 
while the wall model is used to approximate physical quantities
such as wall shear stress or friction, a kinematic shear discontinuity 
(velocity jump tangential to the wall surface) exists.
Different from the Eulerian mesh-based RANS model
where the shear discontinuity does not present in the flow field 
due to the fact that the first computational cell locates at $y_{p}$, 
the Lagrangian particle-based method faces great challenges on 
handling the wall-adjacent unresolved region 
when the standard wall (dummy) particles boundary is used.

In this situation, 
not only the strong shear stress presents the same as mesh-based method, 
the kinematic shear discontinuity presents also 
with the large relative motion between the fluid and wall particles.
Note that such situation is also different from the inviscid WCSPH simulations,
such as that of dambreak flows, where the kinematic shear discontinuity presents
only because the shear stress between fluid and wall particles is neglected 
deliberately to avoid near-wall instabilities. 
Due to the coupling dynamics between the strong shear stress and particle motion,
the small disturbances, such as those introduced by 
the particle migration from the inner region or adverse pressure gradient, 
can amplify, pollute the near-wall particle distribution
and lead to large numerical errors and instabilities.
\subsection{Lagrangian particle-based wall model}
\label{wall-function-sec}
\begin{figure}[tb!]
	\centering
	\includegraphics[trim = 0cm 0cm 0cm 7cm, clip,width = \textwidth]{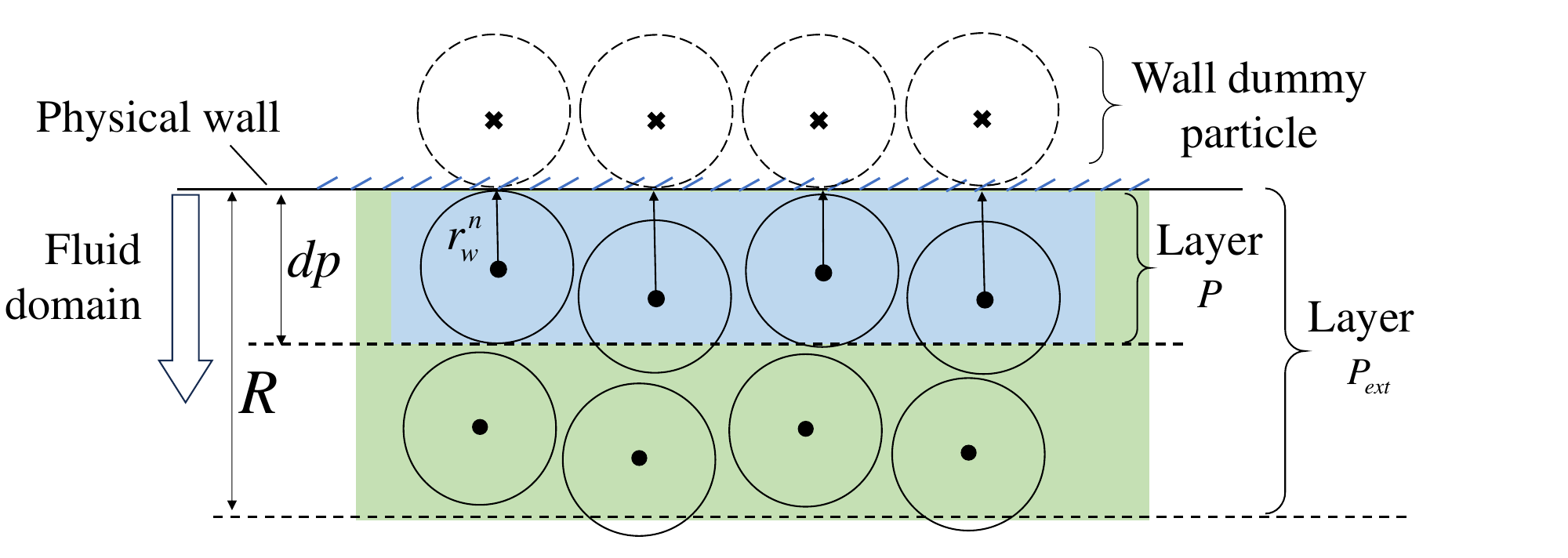}
	\caption{Division of the near wall regions for the particle-based method.
		For all the fluid particles that are near the wall, the distance to wall $r^n_w$ is calculated.
		The particles that satisfy $r^n_w<R$, are defined in the $P_{ext}$ layer, 
		where $R=2h-dp$.
		For those particle, whose $r^n_w<dp$, are defined in the $P$ layer.
	}
	\label{wallfunc-model}
\end{figure}

For the Lagrangian particle-based methods, 
since the near-wall fluid particles move continuously, 
and do not have invariant distance to the surface 
as the wall-adjacent cells in the Eulerian mesh-based methods, 
the wall-adjacent location of 
the original wall model in section \ref{wall-model} 
is handled by two narrow layers,
as illustrated in Figure \ref{wallfunc-model}, 
namely the wall-adjacent and extended layers termed 
as $P$ and $P_{ext}$, where latter covers the former. 
While $P$ layer with the thickness $dp$, where $dp$ is particle spacing, 
is corresponding to the wall-adjacent cells in Eulerian mesh-based method,
$P_{ext}$ layers is introduced due to the fact that the cutoff (influence) 
radius of smoothing kernel, typically $2h$, 
is larger than the thickness of $P$ layer.
Note that, due to  the Lagrangian characteristics, 
the exact distance of the first layer particles to 
the wall surface fluctuates around the average value of $dp/2$ but bounded by $dp$.
For simplicity and numerical stability, 
especially for complex geometric with sharp-corners 
\cite{wang2022simulation},
$y_p = dp/2$ is applied to all the particles within $P_{ext}$ layer.

\begin{figure}[tbh!]
	\centering
	\includegraphics[trim = 0cm 0cm 13cm 4cm, clip,width=0.5\textwidth]{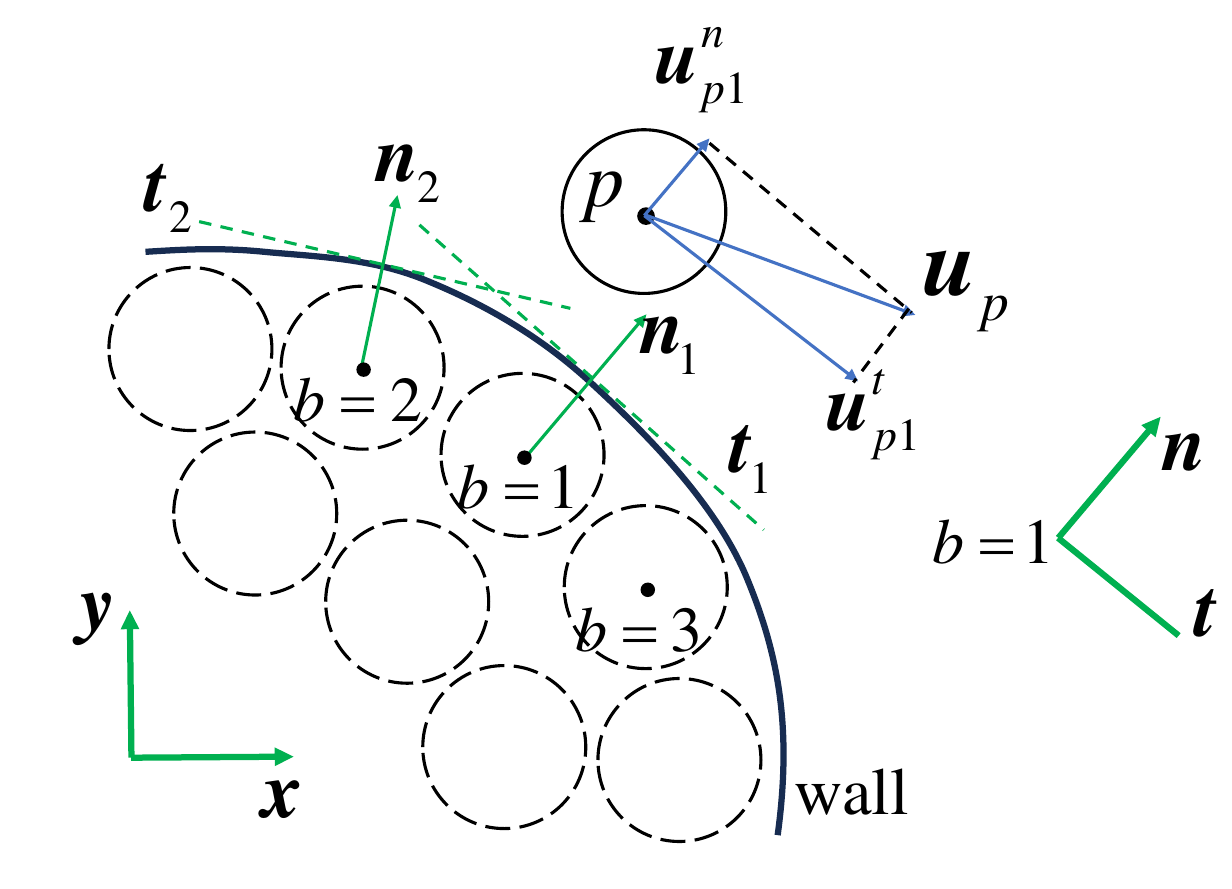}
	\caption{The particle-based wall model for complex geometry when $b=1$,
	$\mathbf{n}_b$ are the unit normal vectors evaluated at wall particles. 
	}
	\label{wallfunc-complex-shape}
\end{figure}

For a particle within $P_{ext}$ layer of a general complex wall surface,
as shown in Figure \ref{wallfunc-complex-shape}, 
one can define the tangential flow velocity as
\begin{equation}
\mathbf{v}^{t}_p =U_p \mathbf{t}_p = \mathbf{v}_p - \mathbf{v}_p \cdot \mathbf{n}_p \mathbf{n}_p,
\label{tangential-velocity}
\end{equation}
where $\mathbf{n}_p =\sum_{b} W_{ib}\mathbf{n}_b/\sum_{b} W_{ib}$ 
is the weighted surface normal evaluated from 
the values $\mathbf{n}_b$ at each neighboring wall particle $b$.
Note that the kernel weight is introduced to include the influence of the complex geometry on tangential velocity.
And hence the tangential flow direction $\mathbf{t}_p$ is computed as
\begin{equation}
\mathbf{t}_p=\frac{\mathbf{v}_p - \mathbf{v}_p \cdot \mathbf{n}_p \mathbf{n}_p}
{|\mathbf{v}_p -\mathbf{v}_p \cdot \mathbf{n}_p \mathbf{n}_p| + \varepsilon},
\label{flow-tangential}
\end{equation}
where $\varepsilon$ is small positive number to avoid dividing zero.
Based on $\mathbf{t}_p$ and Eq. (\ref{eq:wall-shear-stress}), 
the wall shear stress of a particle within the $P_{ext}$ layer
can be expressed as
\begin{equation}
	\bm{\tau}_{w,p}=\frac{\rho U_{p} C^{1/4}_\mu k^{1/2}_p}{u^+} \mathbf{t}_p \otimes \mathbf{t}_p.
	\label{wallfunc-equ-weighted}
\end{equation}
Similarly, the corresponding velocity gradient, required in Eq. \eqref{wallfunc-production-k-equ2} for computing the $G_{k,p}$, is obtained as
\begin{equation}
\nabla\mathbf{v}_{p}=\frac{\partial U_{p}}{\partial n} \mathbf{t}_p \otimes \mathbf{t}_p.
\label{wall-gradient}
\end{equation}
%
\subsection{Wall boundary conditions and weighted compensation scheme}
\label{sec-bc-compensate}
In this work, for the fluid particles near wall or within the $P_{ext}$ layer, 
4 types of wall boundary condition are considered here in respect to 
different RHS terms of the governing equations and the wall model.

The first type is for the divergence term in the mass conservation equation Eq. (\ref{mass-equ})
and the pressure gradient term in the momentum equation Eq. (\ref{momentum2-equ}).
With  dummy wall particles \cite{adami2012generalized},
the one-side Riemann-based boundary condition \cite{zhang2017weakly} is employed.
Note that, due to the large velocity jump between the near-wall particles and wall, 
the low-dissipation limiter used in the original formulation is not employed here
to provide sufficient numerical dissipation.

The second type is for the viscous term 
in the momentum equation Eq. (\ref{momentum2-equ}).
The contribution from wall is computed according to 
the wall shear stress of the wall model as
\begin{equation}
	\left(\frac{\text{d} \mathbf v}{\text{d} t}\right)^\nu _{wall}=  - \frac{2}{\rho} \sum_{b} \boldsymbol{\tau}_{w, p} V_b  \nabla W_{ib}.
	\label{wall-shear-viscous-expand-equ}
\end{equation}

The third type boundary condition is for the diffusion terms 
in the $k$ and $\epsilon$ transport equations, 
i.e. Eqs. (\ref{k-equ}) and (\ref{epsilon-equ}).
It is known that the diffusion terms do not significantly 
affect results\cite{singhal1981predictions}, therefore, 
is neglected assuming the diffusion of $k$ and $\epsilon$ to wall as zero
or adiabatic wall condition.

The fourth type wall boundary condition is for the velocity gradient used 
in the production terms of Eqs. (\ref{k-equ}) and (\ref{epsilon-equ}).
Here, for the particles in $P$ layer, 
the velocity gradient is constrained according to the wall model as Eq. (\ref{production-equ}).
Some complication comes to the particles not in $P$ but $P_{ext}$ layer.
Since the velocity gradient as in Eq. (\ref{wallfunc-production-k-equ2})
is quadratically sensitive to the turbulent kinetic energy 
and very large in this region,
a straightforward application of the non-slip boundary velocity boundary condition
\cite{adami2012generalized} may lead to considerable underestimation 
even with the kernel gradient correction of Eq. \eqref{velo-grad-equ}.

To address this issue, we introduce a different formulation
which only takes account from fluid particles 
but integrating the contribution of the velocity gradient from the wall model,
as that of Eq. (\ref{wall-gradient}), for neighbor particles in $P$ layer as
\begin{equation}
	(\nabla \mathbf{v}_i)^{P_{ext}} =
	\begin{cases}
		\displaystyle\sum_j \mathbf{v}_{ij} \otimes \mathbf{B}_{i}^{in} V_j \nabla W_{ij},                  & {\rm for } ~j \notin P
		\\
		\sum_j \left[w_s \nabla\mathbf{v}_{p,j}
 \mathbf{r}_{ij} +(1-w_s)\mathbf{v}_{ij}\right] 
 \otimes \mathbf{B}_i^{in} V_j \nabla W_{ij}, & {\rm for } ~j \in P
	\end{cases}.
	\label{velo-grad-wall-weight}
\end{equation}
$\mathbf{B}_{i}^{in}$ means the correction matrix that 
is computed according to Eq. \eqref{B-matrix-equ} but only considers the inner neighbors.
Equation \eqref{velo-grad-wall-weight} suggests that, 
when the neighbor particles are in $P$ layer, 
the velocity gradient from the wall function is partially involved to compensate the underestimation.
The weight $w_s$ is fixed at 0.1 when the ARD is deactivated and 0.5 when the ARD is activated in this work.
The effect of this weighted compensation scheme will be tested in section \ref{straight-channel}.
%
\subsection{The constant $y_p$ strategy and boundary-offset technique}\label{sec:refinement}
It is well known that the correctness and accuracy of a numerical method
can be evaluated by convergence study, that is, compare the simulation results 
with increasing resolution to the reference.
Although being straightforward for a general numerical method,
the convergence study would be difficult for a RANS simulation method with the wall model
if $y_p$ is associated with the particles size $dp$, 
as shown in Fig. \ref{wallfunc-model}.
This is because, as the resolution increases , $y_p$ decreases, causing the corresponding $y^+$ falling out (below) the effective range, thereby invalidating the wall model.

To resolve such conflict, we propose a constant-$y_p$ strategy 
so that the rigorous convergence can be carried out.
First, a proper $y_p$ is chosen with the corresponding $y^+$ 
falling well in the effective range.
Then, we set this $y_p$ as a model parameter 
and keep it unchanged with increasing resolution.

\begin{figure}[tb!]
	\centering
	\includegraphics[trim = 0cm 0cm 7cm 4cm, clip,width=0.9\textwidth]{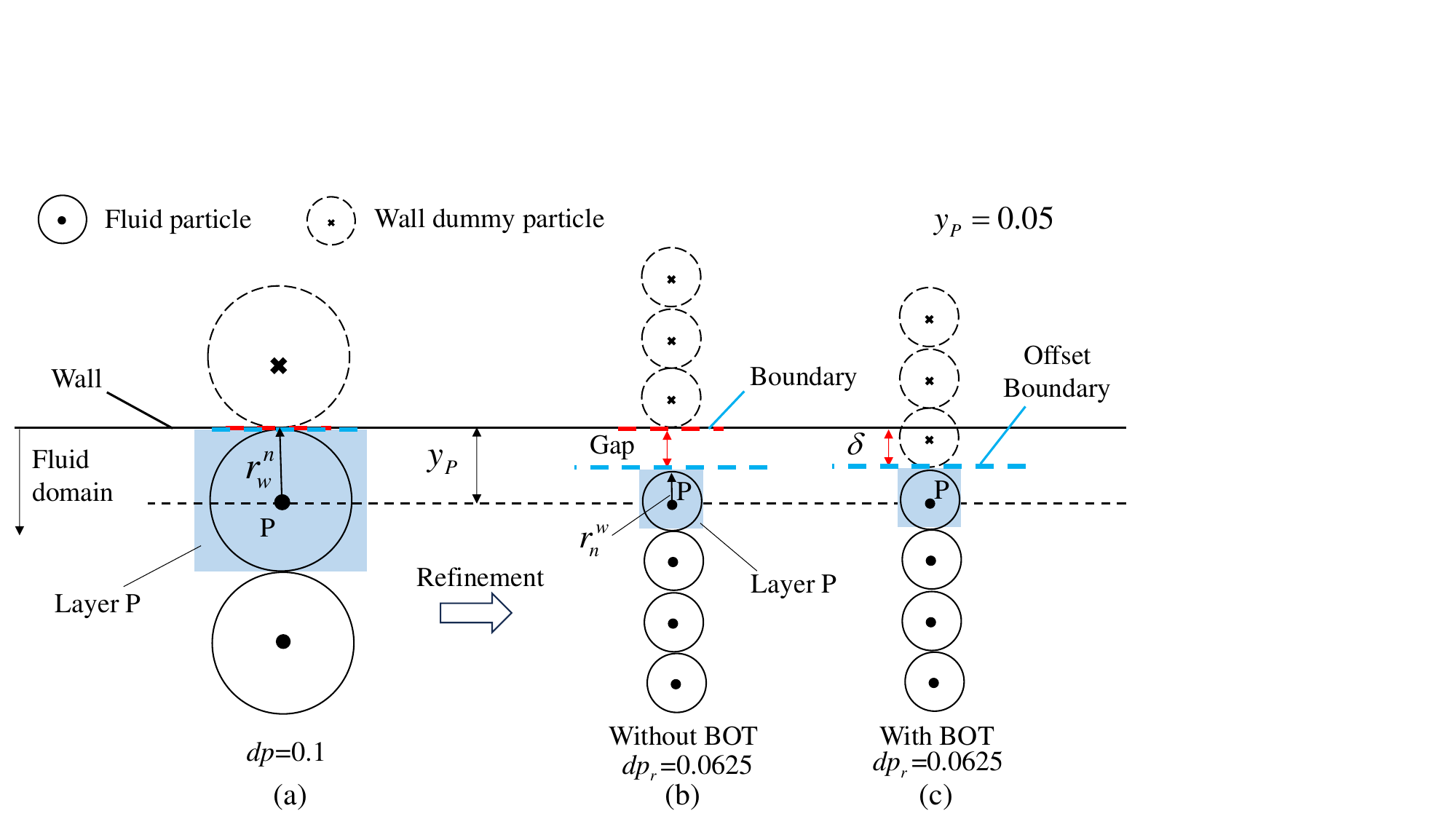}
	\caption{Concept of boundary-offset technique: (a) at the coarse resolution setup, 
		the $P$ layer coalesces with the space between the first layer particles and the wall surface (boundary),
		and the dummy particles are used to represent the original solid wall;
		(b) at a refinement setup, the thickness of $P$ layer is decreased to $dp_r$, 
		i.e. the refined particle spacing, but a gap appears if not use boundary-offset technique(BOT); 
		(c) with the BOT, the boundary is offset with a distance $\delta = y_p - dp_r/2$, 
		and the dummy particles represents the extended wall.}
	\label{offset-model}
\end{figure}

However, the constant-$y_p$ strategy introduces another issue 
on handling the wall boundary. 
With the constant $y_p$ and decreasing particles size $dp$,
a gap will be generated between the $P$ layer and the wall dummy particles, as shown in Fig. \ref{offset-model} (b).
To handle this issue, a boundary-offset technique (BOT) is developed.

As demonstrated in Figure \ref{offset-model}, 
the basic idea is to offset the boundary 
so that the gap between the first layer, or $P$ layer particles, and
the original wall is filled.
The offset boundary technique is consistent with the wall model, because the wall boundary values, such as the wall shear stress, are actually imposed on the first-layer particles according to the wall function. 
Note that, with increasing resolution,
both the fluctuations of the exact distance to 
the wall surface of the first layer particles
and the difference between $P_{ext}$ and $P$ layers vanish,
indicating more accurate numerical representation 
of the wall model.

\begin{figure}[tb!]
	\centering
	\includegraphics[trim = 0cm 0cm 17cm 2cm, clip,width=0.70\textwidth]{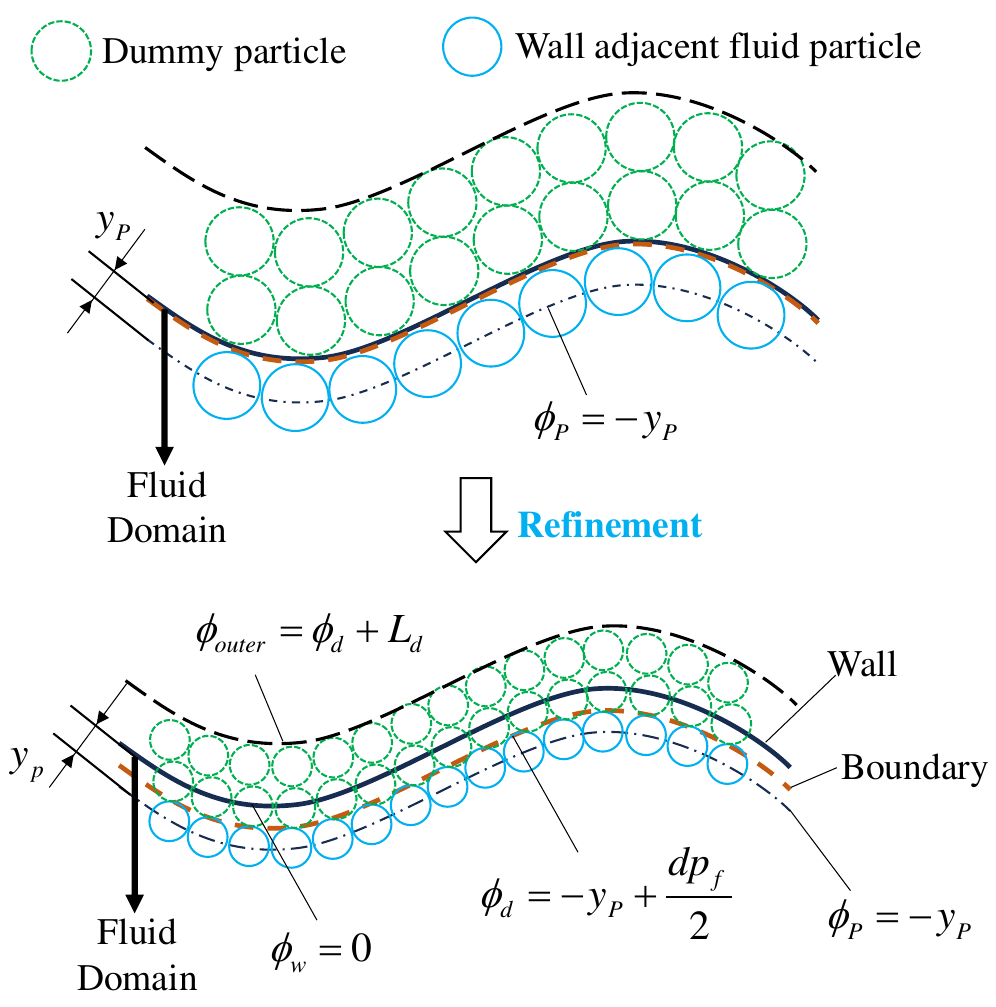}
	\caption{Boundary-offset technique for complex geometry.  
	$\phi_w=0$ represents the wall,
	 $\phi_d=-y_p+dp_f/2$ refers to the boundary (interface between the fluid and wall dummy particles), and $dp_f$ is the fluid particle diameter.
	$\phi_{outer}=\phi_d+L_d $ is the outer bound of dummy particles, where $L_d = 4dp_d$ is the thickness of the dummy particle region, and $dp_d$ is the dummy particle diameter.
}
	\label{Offset-model-complex}
\end{figure}

The implementation of the BOT for simple geometry is straightforward,
however, it may lead to complication for more general complex geometry. 
In the present work, the level set function is used to describe a complex geometry\cite{zhu2021cad} and achieve the offset operation. 
As shown in Fig. \ref{Offset-model-complex}, 
the offset boundary can be obtained by extruding 
the original zero level set function (representing the solid wall surface) with the offset distance.
Consequently, the dummy particles are generated in the 
region between $\phi_d$ and $\phi_{outer}$, and the initial fluid particles are generated inside the boundary $\phi_d$.
For both fluid and dummy particles, initial relaxation is necessary to increase start-up stability \cite{zhu2021cad}.
\begin{figure}[htb!]
	\centering
	\includegraphics[trim = 0cm 0cm 15cm 12cm, clip,width=0.80\textwidth]{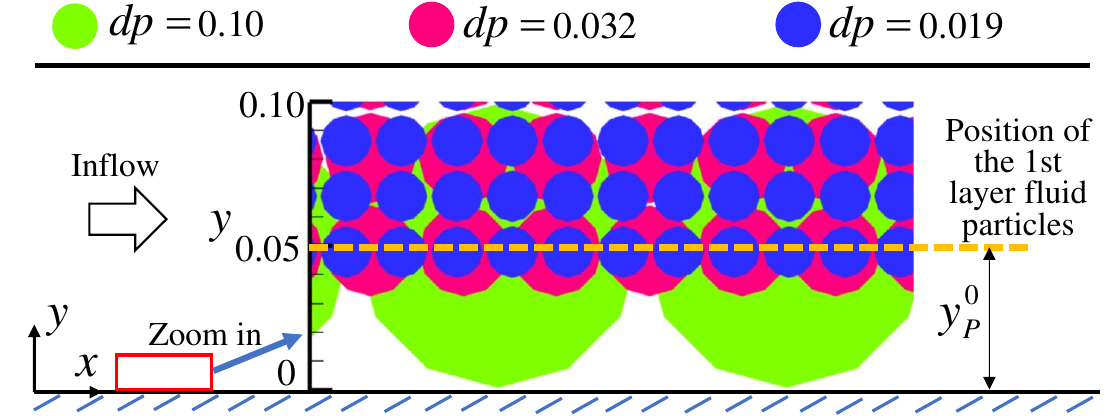}
	\caption{The initial fluid domain for straight channel under different resolutions when use the offset model.}
	\label{straight_cha_fluid0}
\end{figure}

\begin{figure}[htb!]
	\centering
	\includegraphics[trim = 0cm 0cm 9cm 8cm, clip,width=0.80\textwidth]{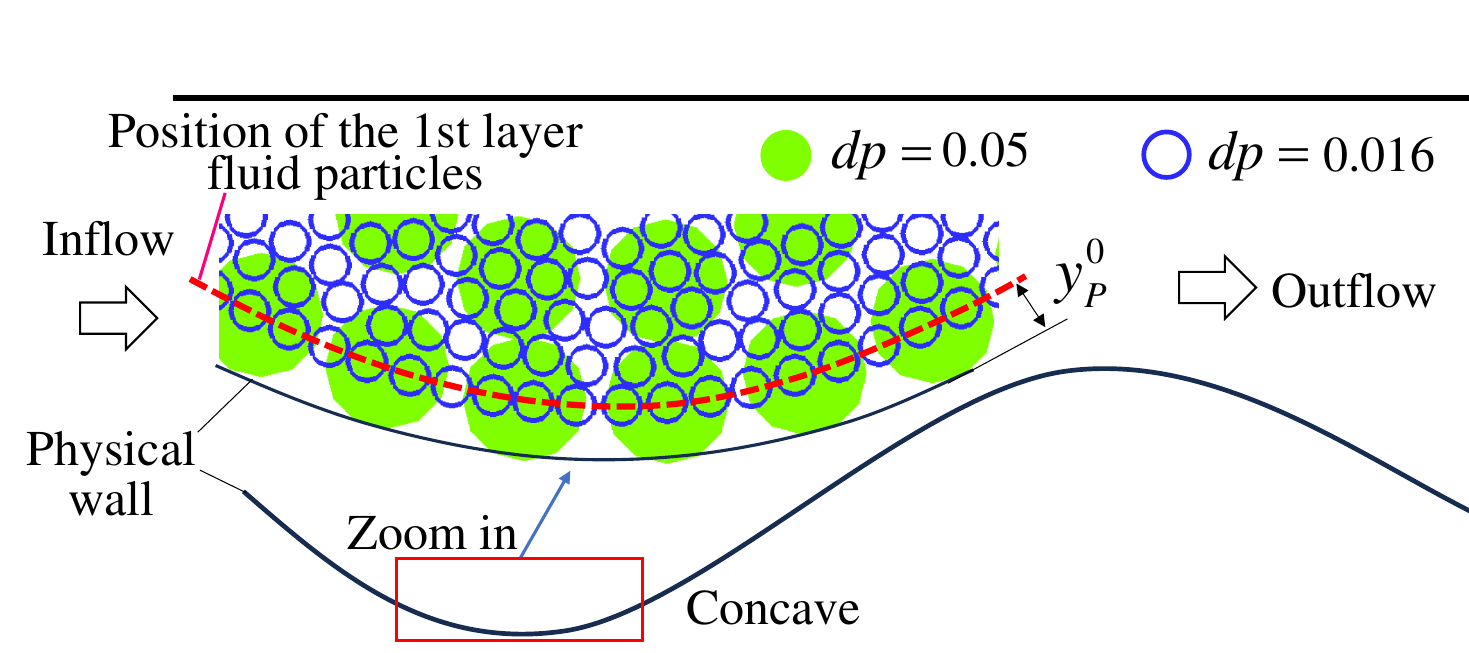}
	\caption{The particle distribution for wavy channel under the two different resolutions when use the offset model.}
	\label{wavy_cha_fluid0_dummy0}
\end{figure}

Figure \ref{straight_cha_fluid0} shows the initial fluid particles under different resolutions for the straight channel case.
With the resolution increasing, the distance to wall of the first layer particle is the same, meaning that the $y^+$ remains consistent.
Figure \ref{wavy_cha_fluid0_dummy0} shows the distribution of both fluid and dummy particles at the initial stage under two different resolutions for the wavy channel shape.
The distance from the first layer fluid particles to physical wall is the same under different resolutions for this complex shape.

It is worthy noting that the boundary-offset technique is used 
for evaluating the correctness and accuracy of a present numerical method
when FSI is not considered.
When FSI is considered, since the solid dynamics needs to be solved
within the original wall surface, the boundary-offset technique is not applicable anymore.
In that case, the resolution of the flow simulation is constrained by the wall model.
To achieve a rigorous convergence study for the entire FSI system, 
one possible solution is using more advanced wall models 
which are independent on the value of $y^+$.
%
\section{Numeral examples}
\label{section-numerical-examples}
In this section, we first consider the fully-developed turbulent straight channel case to comprehensively verify the proposed method and test the effect of the present schemes.
Second, the mildly- and strongly-curved turbulent channels are calculated for testing the method within the non-orthogonal coordinate.
Last but not least, the half-converging and diverging turbulent channel case, which involves flow separation, is simulated to demonstrate both the aforementioned inconsistency and the capability of the proposed method to address it. 
Again, the C2 Wendland kernel is utilized, with the smoothing length set to be $h = 1.3dp$ if not stated otherwise.
Also note that convergence studies are carried out for all test cases by employ the boundary-offset technique.
\subsection{Fully-developed turbulent channel flow}
\label{straight-channel}
The fully-developed turbulent channel flow is a typical benchmark case for testing 
numerical methods for wall-bounded turbulent flow simulation.
This case has been commonly used to verify different methods, 
such as DNS\cite{d2004numerical,moser1999direct}, 
LES\cite{moin1982numerical,maeyama2023near} 
and RANS\cite{zhou2017assessment,chien1982predictions}, 
while most of them are mesh-based.
To make a reliable and rigorous comparison, 
firstly we simplified this case to the one-dimensional steady problem.
For the fully-developed channel flow, the governing equations can be simplified as\cite{hussain1975measurements},
\begin{equation}
	\begin{cases}
		(\nu + \nu_t) \frac{d \bar{u}}{d y} = u_{\tau}^2 \left(1 - \frac{y}{H/2}\right)                                                                                                \\
		\nu_t \left(\frac{d \bar{u}}{d y}\right)^2 - \epsilon + \left(\nu + \frac{\nu_t}{\sigma_k}\right) \frac{d^2 k}{d y^2}=0                                                        \\
		C_1 \frac{\epsilon}{k} \nu_t \left(\frac{d \bar{u}}{d y}\right)^2 - C_2 \frac{\epsilon^2}{k} + \left(\nu + \frac{\nu_t}{\sigma_{\epsilon}}\right) \frac{d^2 \epsilon}{d y^2}=0 \\
	\end{cases},
	\label{ODE_straight_channel}
\end{equation}
where $u_\tau$ is the friction velocity, $H$ is the channel height 
and $y$ refers to the normal distance from wall.
With the three unknown variables and three equations, 
theoretically, this Ordinary Differential Equation (ODE) system can be solved 
by employing Finite Difference Method (FDM).
Specifically, the central difference scheme is used for the second-order differential operators, 
and the velocity gradient is discretized by the backward scheme.
Since the source terms of the $k$ and $\epsilon$ equations are highly non-linear, 
the source term linearization scheme is used.
The Reynolds number is 40000 based on the average mean velocity and the channel height.
Other parameters are the same as the DNS
simulation\cite{lee2015direct} where the friction Reynolds number is 543.496.
The calculation is based on Python 3.12, 
and all the parameters are set the same as the SPH simulation.

\begin{figure}[tb!]
	\centering
	\includegraphics[trim = 0cm 0cm 14cm 3cm, clip,width=0.85\textwidth]{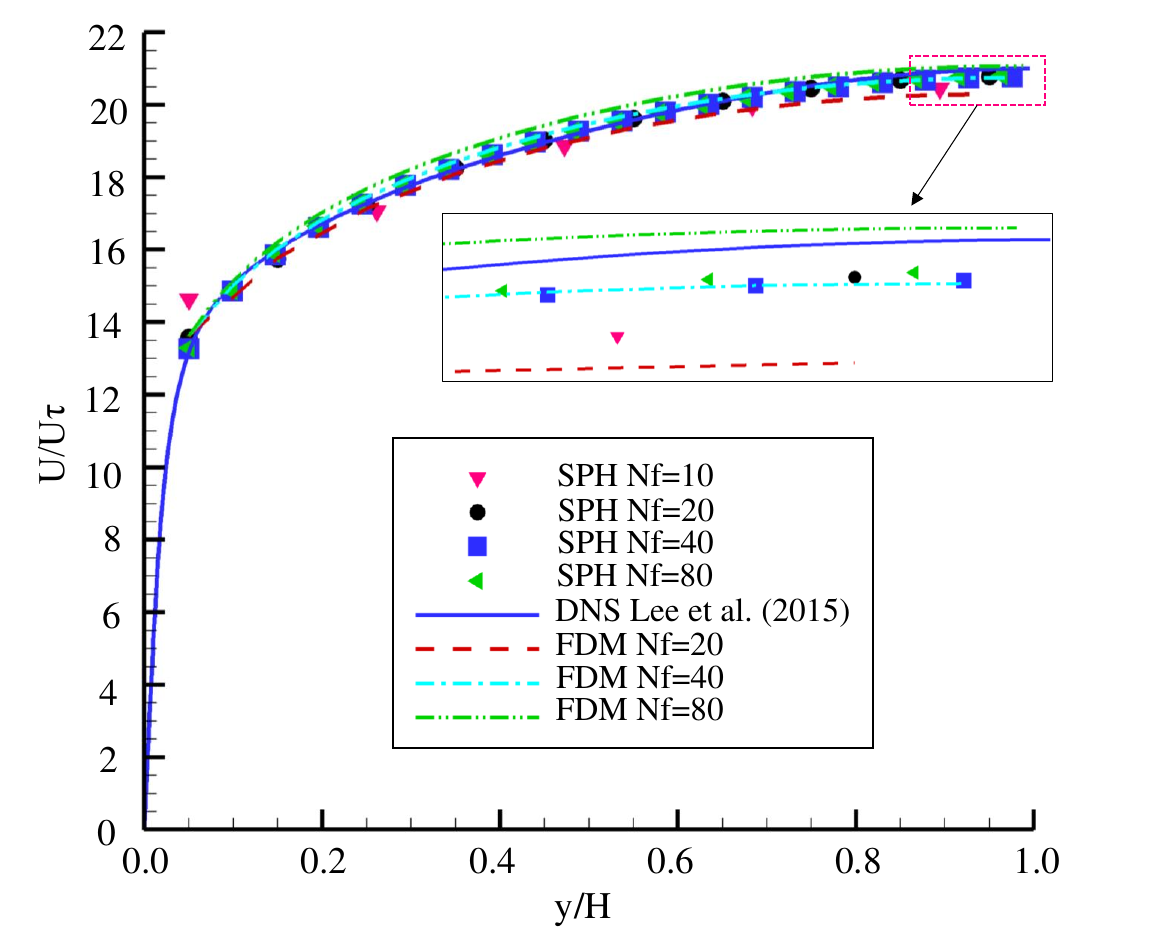}
	\caption{Velocity convergence profiles in a linear scale for the FDM and SPH method.}
	\label{straight_chan_convergence-vel-linear}
\end{figure}

\begin{figure}[htb!]
	\centering
	\includegraphics[trim = 0cm 0cm 16cm 2cm, clip,width=0.8\textwidth]{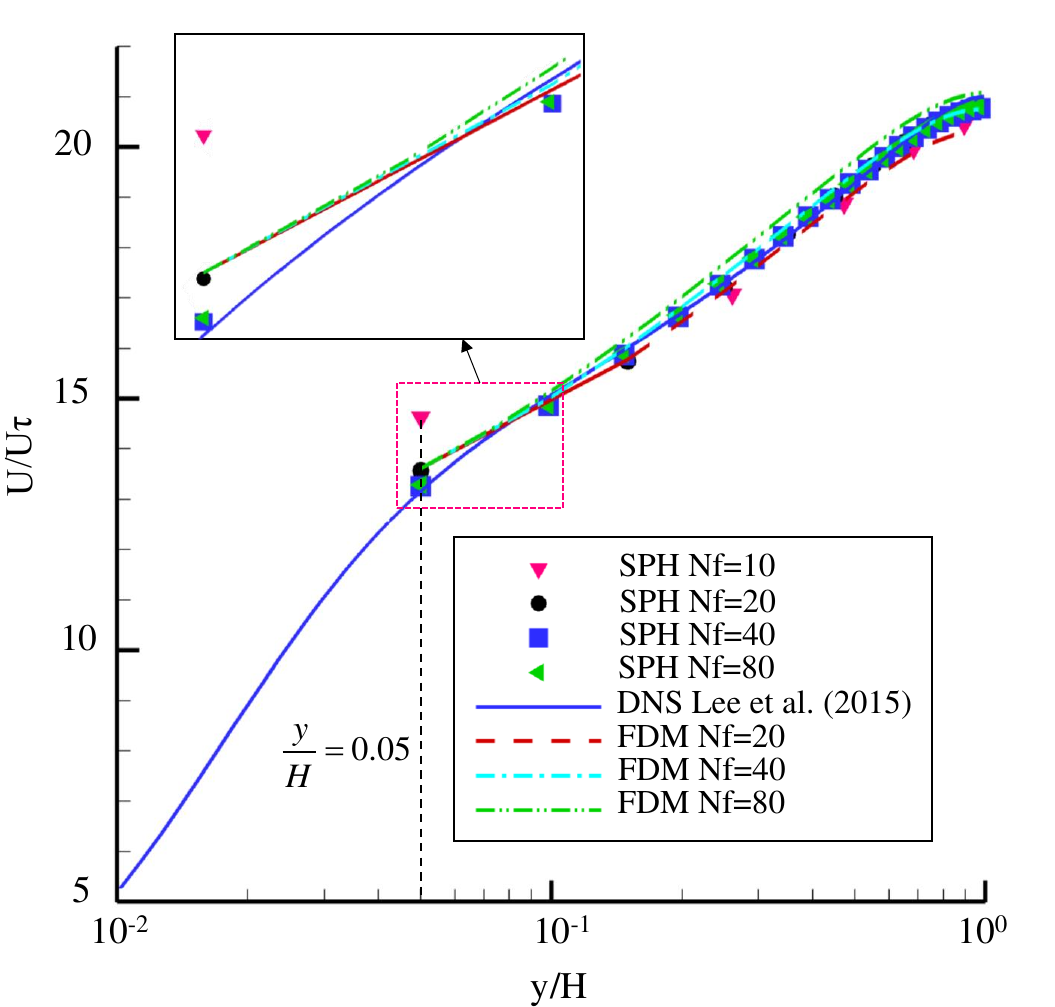}
	\caption{Velocity convergence profiles in a logarithmic scale for the FDM and SPH method.}
	\label{straight_chan_convergence-vel-log}
\end{figure}

\begin{figure}[htb!]
	\centering
	\includegraphics[trim = 0cm 0cm 11cm 4cm, clip,width=0.9\textwidth]{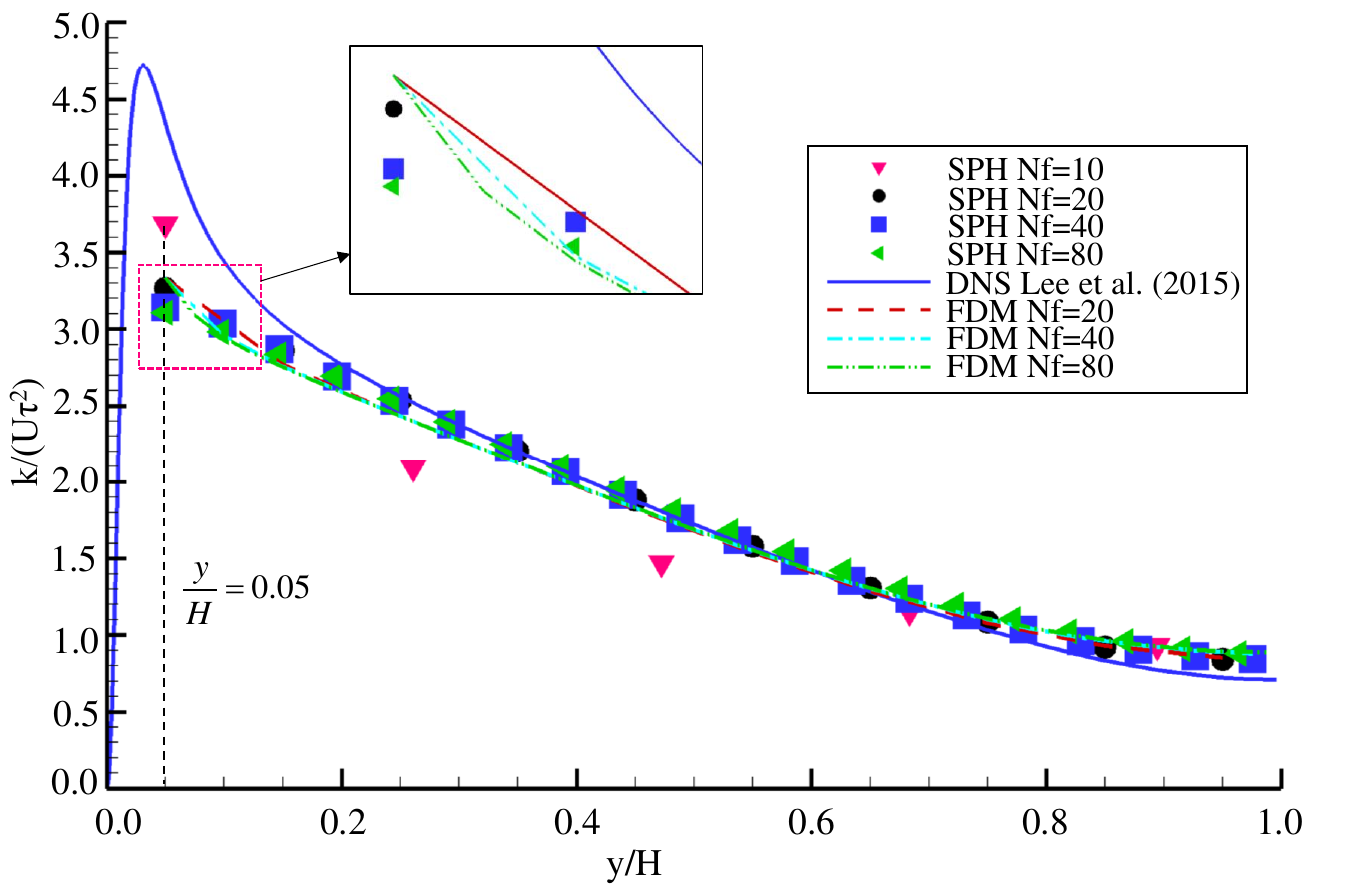}
	\caption{Turbulent kinetic energy convergence profiles for the FDM and SPH method.}
	\label{straight_chan_convergence_k}
\end{figure}

For the SPH simulation, the velocity inflow and zero pressure outflow boundary conditions are used \cite{zhang2024dynamical}. 
The data from the FDM result is used as the input data to accelerate the flow development.
Since there is no strong flow separation in this case, the ARD technique is not activated for the convergence test.
The velocity profiles at the outlet cross section calculated by the FDM and SPH methods under different resolutions are shown in Fig \ref{straight_chan_convergence-vel-linear} (linear-coordinate) and \ref{straight_chan_convergence-vel-log} (log-coordinate).
Note that, time average is conducted for all the quantitative data of the SPH method.
The mean flow velocity is normalized by the friction velocity.
And $N_f$ means the number of the fluid particles on the cross-section.
Both the SPH and FDM results show good convergence.
The SPH result is close to the DNS result when $N_f$ is 20 and the maximum difference  is 3.1\%, and this difference decreases to 0.75\% when $N_f=40$.
Figure \ref{straight_chan_convergence_k} shows the normalized turbulent kinetic energy on the cross-section for different resolutions.
By introducing the limited TVF, the $k$ value also converges.
The $k$ over-prediction problem for the SPH-RANS implementation \cite{bao2023pof,wang2022isph,violeau2007numerical} is addressed.

\begin{figure}[htb!]
	\centering
	\includegraphics[trim = 0cm 0cm 16cm 4cm, clip,width=0.8\textwidth]{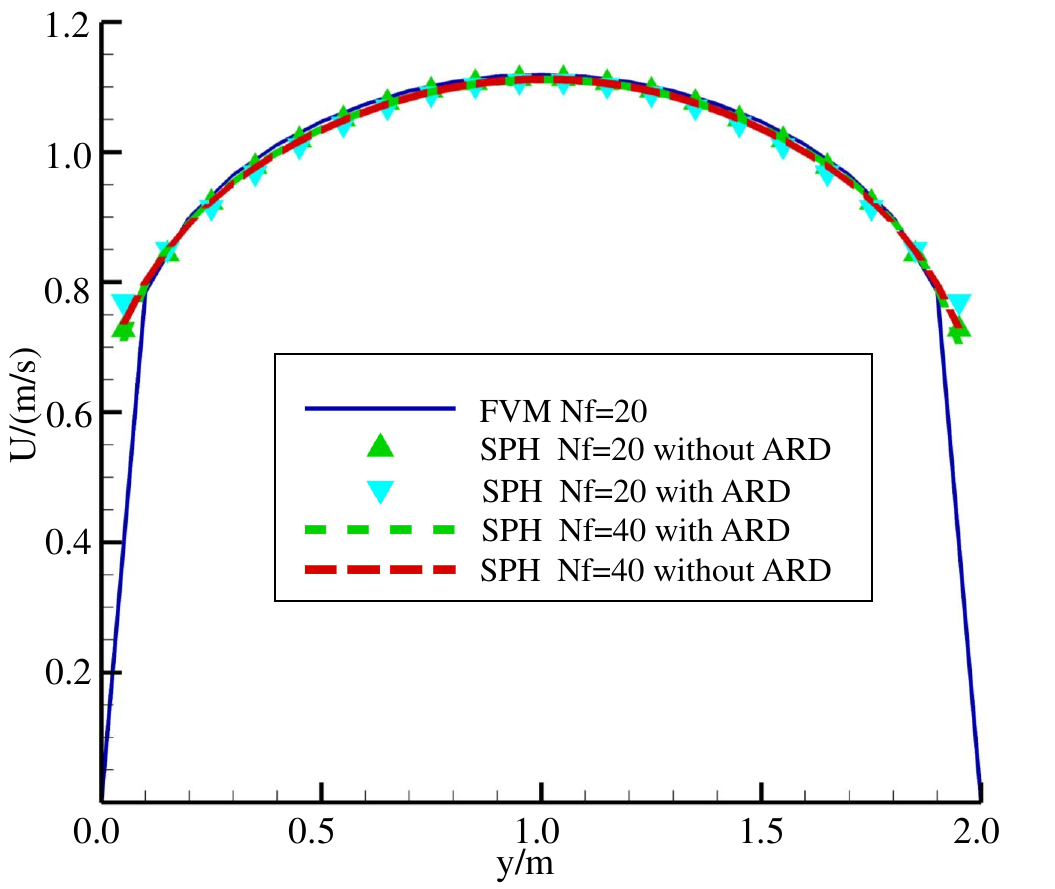}
	\caption{Comparison with the FVM result on mainstream velocity and the effect of the proposed technique.}
	\label{straight_chan_model_effect}
\end{figure}

Figure \ref{straight_chan_model_effect} demonstrates the effect of the proposed techniques, 
including ARD and weighted velocity gradient compensation,  on the velocity and $k$ profiles.
Activation of the ARD technique makes the particle distribution very lattice and hence the degree of kernel truncation is changed.
That is the reason why the weight $w_s$ is recalibrate to 0.5 when the ARD is used.
The results are also compared with the FVM ones, and the FVM simulation is carried out by OpenFOAM-v1912 with all the same parameter settings.

\begin{figure}[htb!]
	\centering
	\includegraphics[trim = 0cm 0cm 16cm 4cm, clip,width=0.8\textwidth]{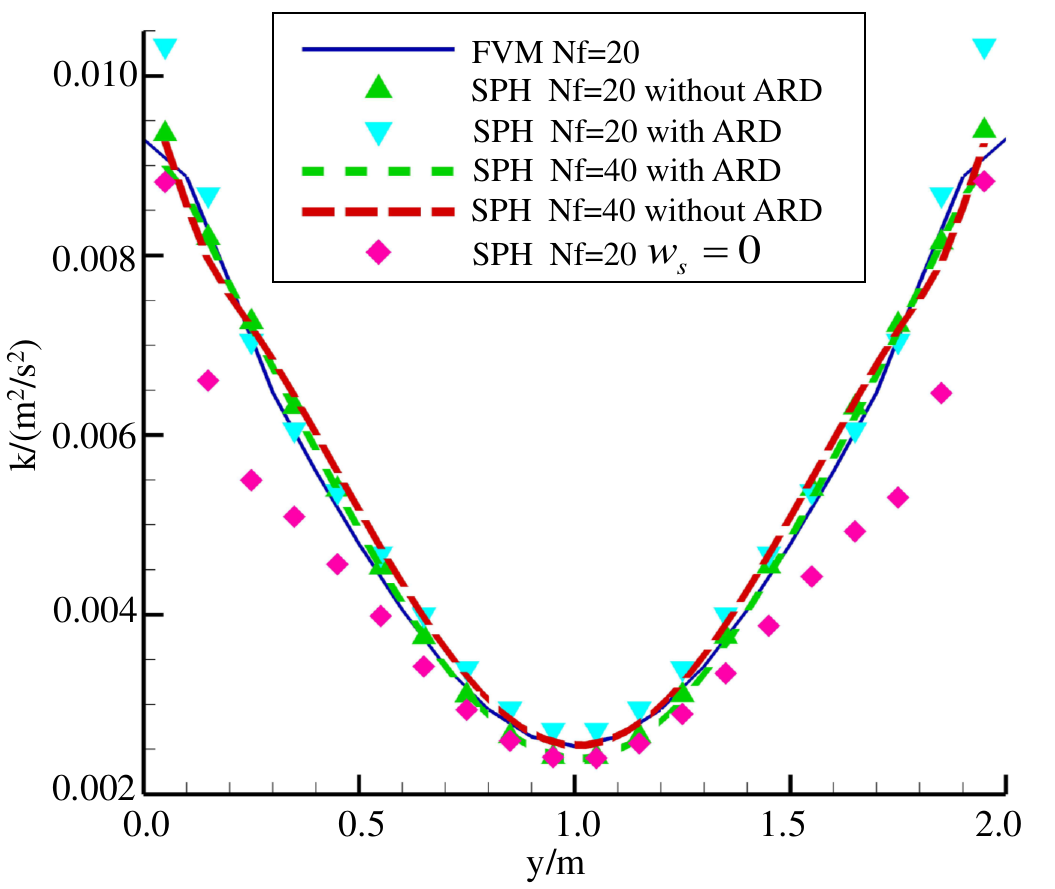}
	\caption{Comparison with the FVM result on turbulent kinetic energy and the effect of the proposed technique.}
	\label{straight_chan_model_effect_k}
\end{figure}
Both the SPH and FVM results agree well with the DNS data at the two resolutions.
And introducing the ARD technique causes minor influence on the result.
Only the wall-nearest velocity and turbulent kinetic energy are slightly bigger when $N_f=20$, but this difference will disappear when the resolution increases to $N_f=40$.

\begin{figure}[htb!]
	\centering
	\includegraphics[trim = 0cm 0cm 14cm 3cm, clip,width=0.8\textwidth]{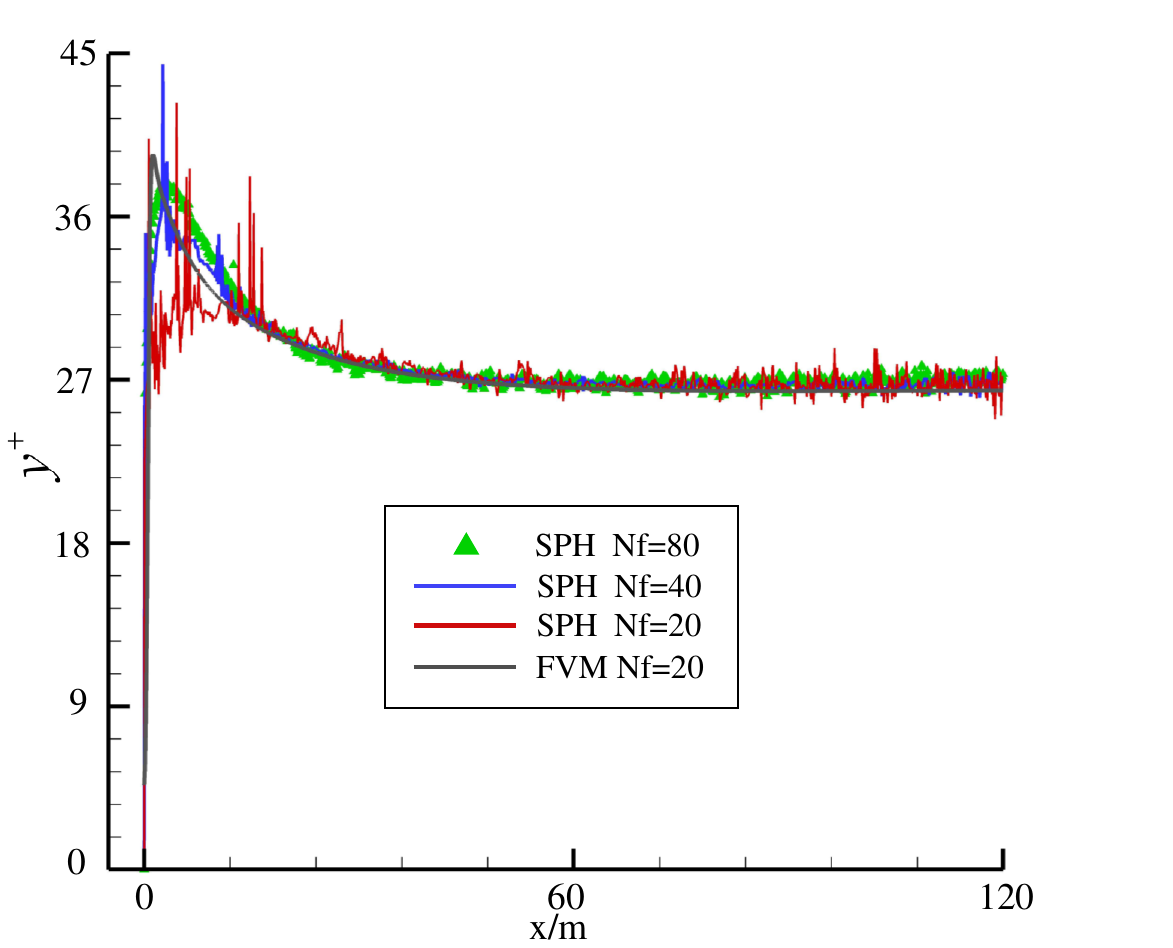}
	\caption{The $y^+$ on the wall-adjacent particles at different resolutions.}
	\label{yplus-at-different-resolutions}
\end{figure}

As for the weighted compensation for the velocity gradient, without this scheme, 
the sub-wall-nearest $k$ is significantly smaller although with the $B_{inner}$ compensation.
To show the effect of the boundary-offset technique, 
the $y^+$ on the wall-adjacent particles under different resolutions is presented in Fig. \ref{yplus-at-different-resolutions}.
The increasing resolution does not affect the $y^+$, proving the effectiveness and rigorousness of the convergence test.
The time-average $y^+$ of the SPH method is 26.69, which is 1.1\% higher than the one calculated by the FVM (26.40).
The friction velocity is 0.0536 m/s, merely 0.94\% higher than that calculated by the finite volume method(0.0530 m/s).
%

\subsection{Turbulent fully-developed curved channels}

\begin{figure}[htb!]
	\centering
	\includegraphics[trim = 0.0cm 0cm 7cm 4cm, clip,width=1.0\textwidth]{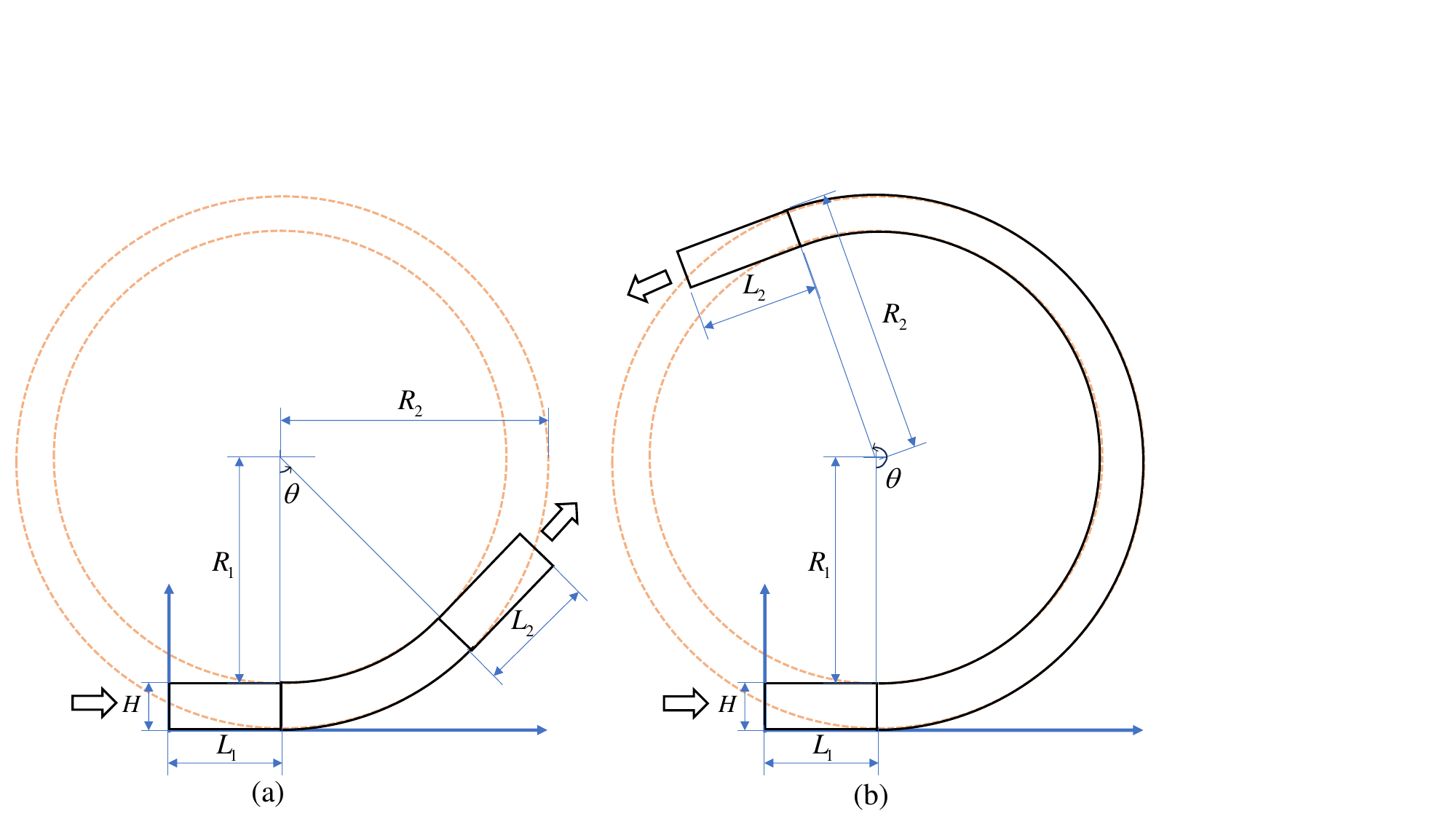}
	\caption{Geometry of the (a) mildly-curved and (b) strongly-curved channels.}
	\label{geo-curved-channels}
\end{figure}

Investigation on the turbulent flow along curved wall is 
of importance for studying the flows in complex fluid machinery.
And many experiments\cite{hunt1979effects,eskinazi1956investigation,ellis1974turbulent} and simulations\cite{brethouwer2022turbulent,li2022piv,pourahmadi1983prediction} have been conducted on this aspect.
In this section, the mildly-curved and strongly-curved turbulent channels 
are simulated to verify the present method in the non-orthogonal coordinate.
The geometry is shown in Fig. \ref{geo-curved-channels}, and the corresponding parameters are listed in Table \ref{geo-curved-channels-para}.
The velocity inlet and zero pressure outlet boundary conditions are used, and the Reynolds number for the two channels are 60000 and 148400, respectively, the same as the experiments\cite{hunt1979effects,eskinazi1956investigation}.
\begin{table}
	\scriptsize
	\centering
	\caption{Geometrical parameters for the curved channels.}
	\begin{tabularx}{8.5cm}{@{\extracolsep{\fill}}lcc}
		\hline
		Parameters                           & Mildly-curved & Strongly-curved \\
		\hline
		Channel height $H$ (m)               & $0.0635$      & $0.0762$        \\
		\hline
		Curvature ratio                      & $100$         & $10$            \\
		\hline
		Inner radius $R_1/H$                 & $99.5$        & $9$             \\
		\hline
		Outer radius $R_2/H$                 & $100.5$       & $10$            \\
		\hline
		Inlet channel length $L_1/H$         & $2.36$        & $3$             \\
		\hline
		Outlet channel length $L_2/H$        & $2.36$        & $9$             \\
		\hline
		Central angle $\theta$ (\textdegree) & $43$          & $210$           \\
		\hline
	\end{tabularx}
	\label{geo-curved-channels-para}
\end{table}

\begin{figure}[htb!]
	\centering
	\includegraphics[trim = 1cm 0cm 1cm 8cm, clip,width=1.0\textwidth]{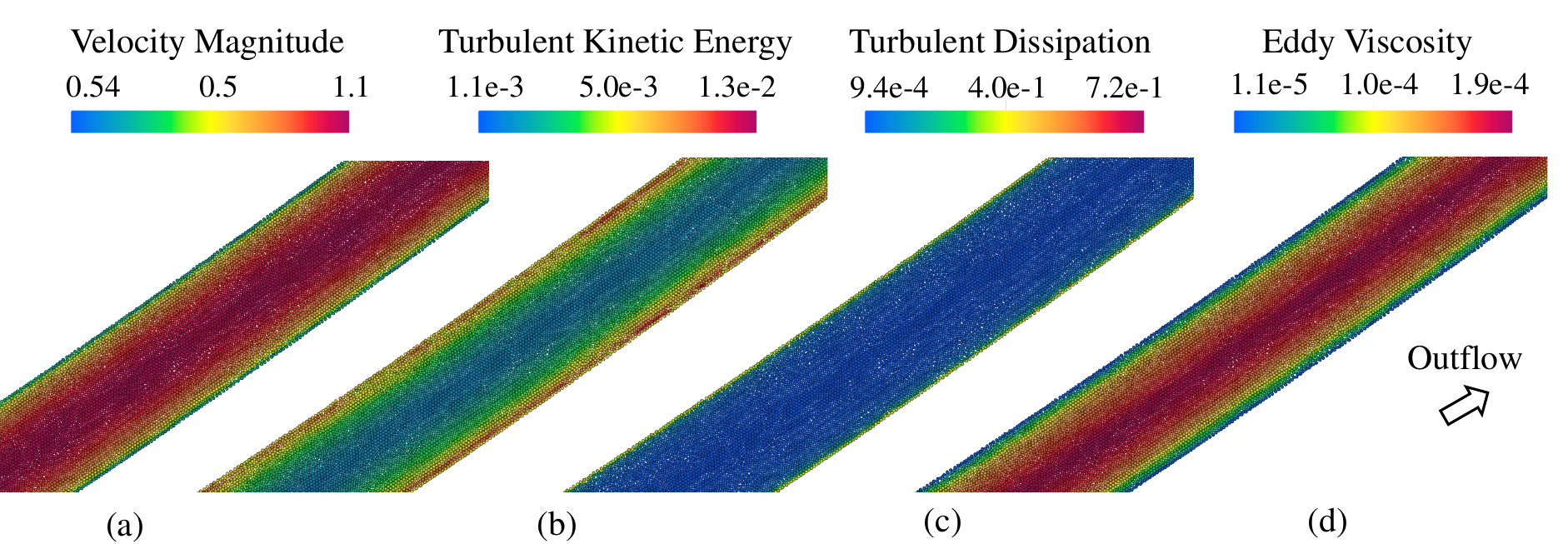}
	\caption{Contours of the mildly-curved channel near outlet.}
	\label{mildly-curved-channels-results}
\end{figure}

\begin{figure}[htb!]
	\centering
	\includegraphics[trim = 0cm 0cm 13cm 3cm, clip,width=0.7\textwidth]{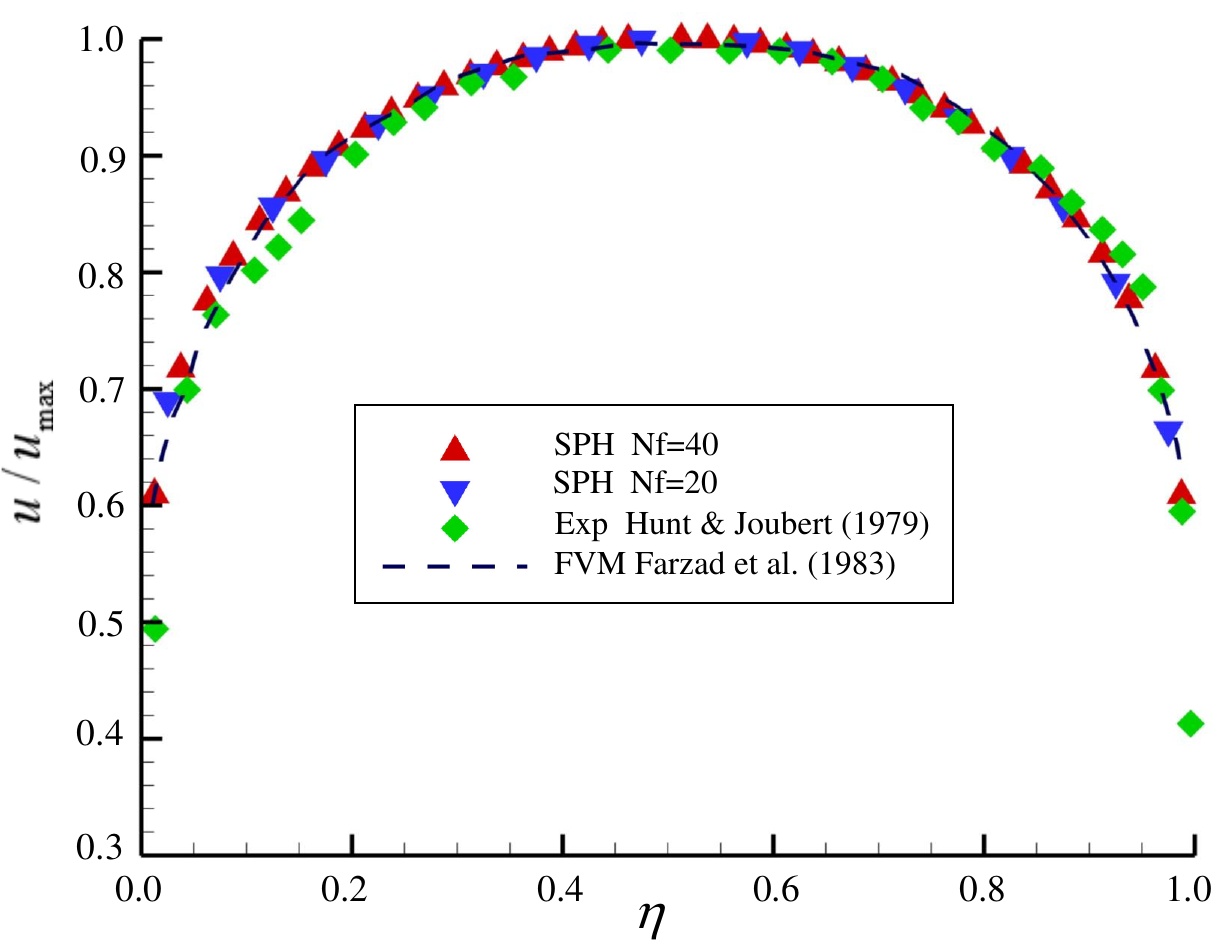}
	\caption{Outlet velocity comparison for the mildly-curved channel.}
	\label{mildly-curved-channels-results-vel}
\end{figure}

For the mildly-curved channel, the contours of the velocity and the three turbulent variables are represented in Fig. \ref{mildly-curved-channels-results}.
The quantitative comparison of the outlet velocity 
is shown in Fig. \ref{mildly-curved-channels-results-vel}.
The SPH results at the two different resolutions agree well with the experiment \cite{hunt1979effects} and the FVM result.
For this simulation, we test the two different $y_p$ values: $y_p=1.59 \times 10^{-3}$ for $N_f=20$ and $y_p=7. \times 10^{-4}$ for $N_f=40$ .
The velocity profiles calculated by the two situations almost overlap with each other and reducing $y_p$ merely improve the near wall resolvability.


For the strongly-curved case, first of all, the flow field initialization could be a critical problem since the WCSPH method is sensitive to the start-up impulse from the velocity inlet.
The common technique includes using a gradually increasing inlet velocity, together with a time-decay acceleration or inlet pressure to accelerate the particles from the zero velocity.
This treatment is effective for straight \cite{wang2022simulation} and mildly-curved channels.
But it would encounter issues when the central angle of the curve channel is more than 180 degree since particles near the outlet will be wrongly motivated.
Besides, tuning the initial turbulent quantities can also be useful\cite{wang2022isph}, but this remedy makes the simulation highly sensitive to the initial conditions.

To achieve a stable start-up without initial condition reliance, we develop a delayed activation scheme, specifically for the Lagrangian turbulence simulation.
That means initially the turbulent module is not activated except from the calculation of the turbulent viscous force. And the flow could be fully laminar by introducing a fixed high eddy viscosity ( $10^{-3}$ in this case).
When the laminar velocity field becomes stable, the turbulent module is then activated.
This technique makes the simulation insensitive to the initial conditions and increases the simulation stability considerably.

\begin{figure}[htb!]
	\centering
	\includegraphics[trim = 0cm 0cm 0cm 5cm, clip,width=1.0\textwidth]{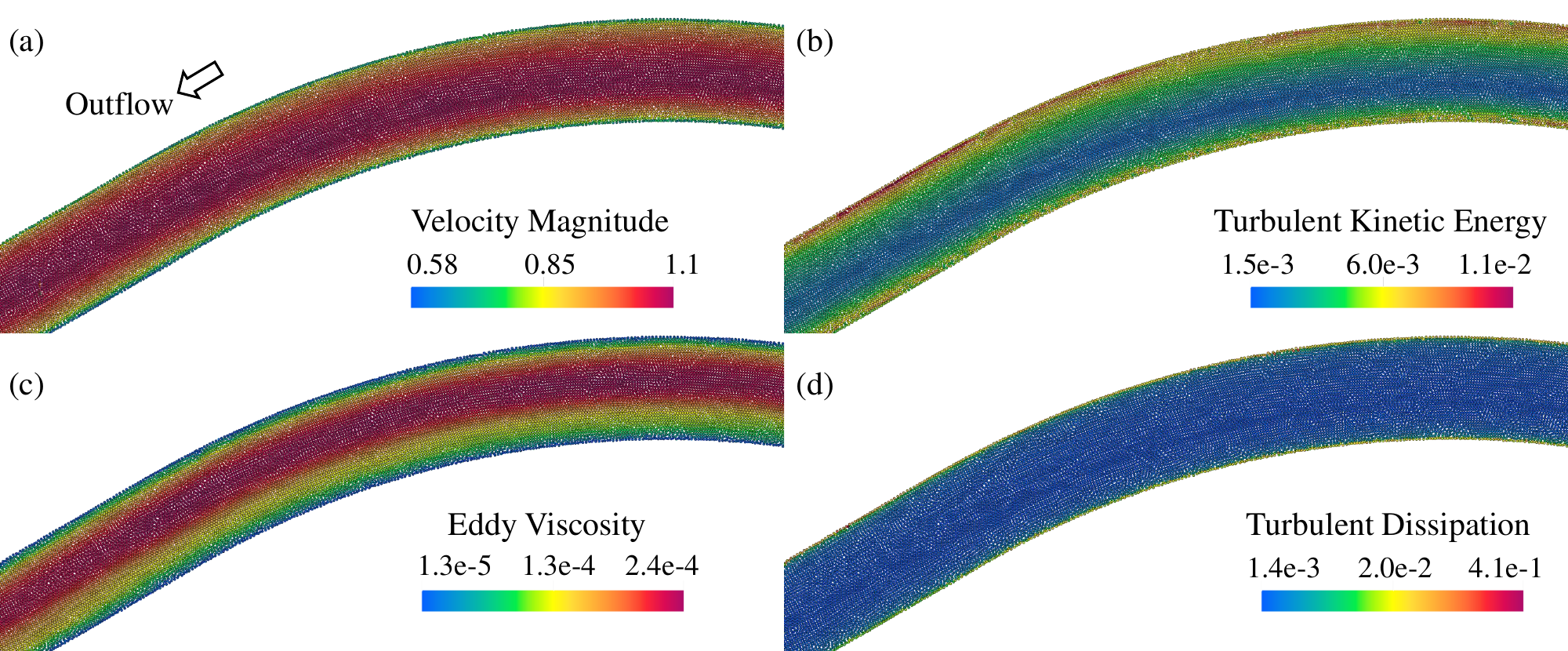}
	\caption{Contours of the strongly-curved channel near outlet.}
	\label{strongly-curved-channel-2D}
\end{figure}

\begin{figure}[htb!]
	\centering
	\includegraphics[trim = 0cm 0cm 11cm 3cm, clip,width=1.0\textwidth]{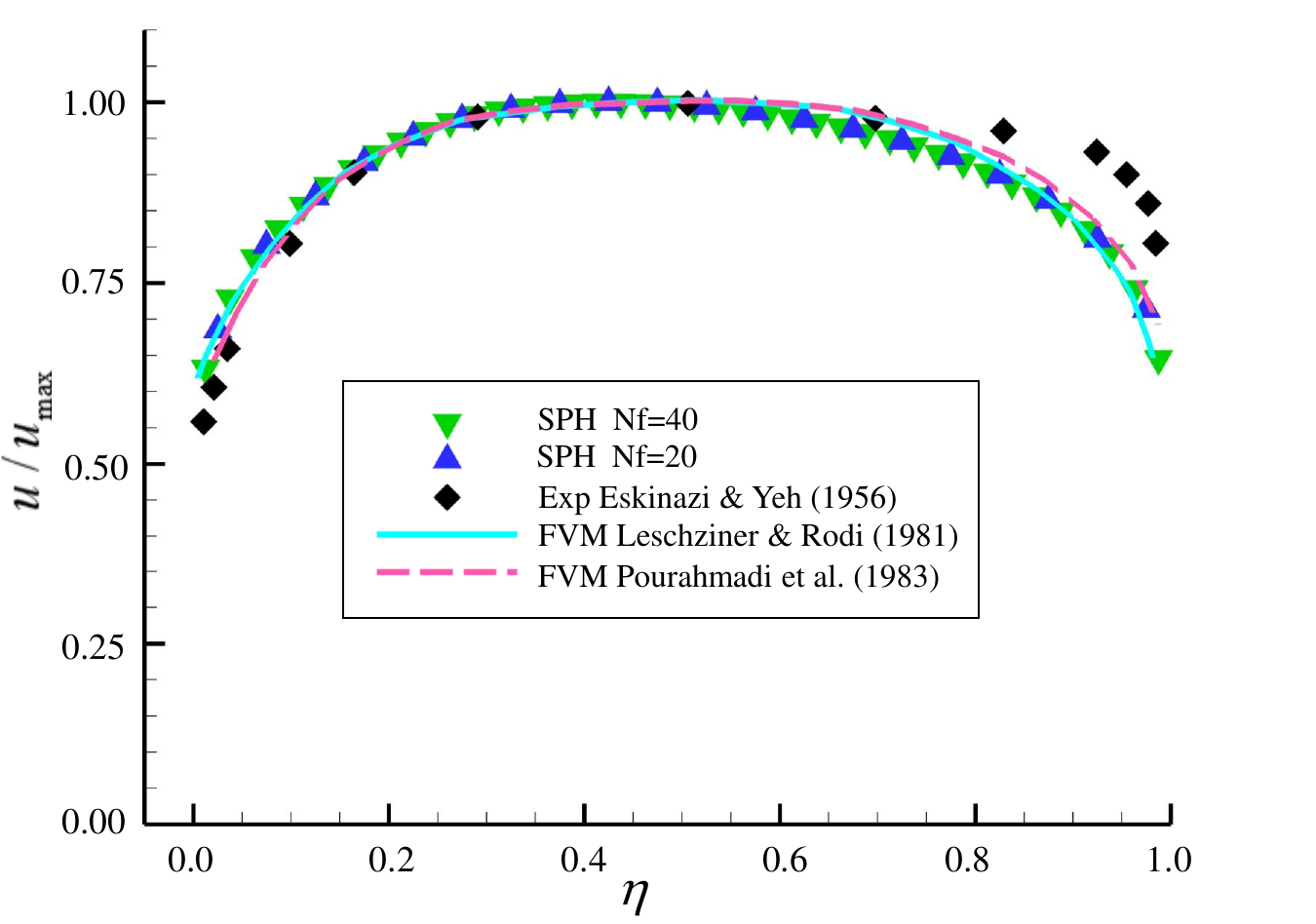}
	\caption{Comparison of the outlet velocity for the strongly-curved channel.}
	\label{strongly-curved-channel-vel-comp}
\end{figure}

The contours of the velocity and three turbulent variables near the curve outlet are shown in Fig. \ref{strongly-curved-channel-2D}.
All of the four variables achieve the fully-developed state.
Figure \ref{strongly-curved-channel-vel-comp} shows the comparison of the velocity and turbulent kinetic energy, $k$, profiles calculated by the proposed method, FVM\cite{pourahmadi1983prediction}, and from the experiment\cite{eskinazi1956investigation}.
The SPH simulations are conducted at the two different resolutions, and a very good convergence is observed for both velocity and $k$.
The FVM results are obtained by using the standard and extended $k-\epsilon$ models\cite{pourahmadi1983prediction}. 
As for the velocity profile, all the numerical methods can achieve an good agreement.
And the maximum difference between the SPH method and the FVM is merely 2.4\%.
Although, when compared with the experiment, all of them under-predict the velocity profile near the outer curved wall because of the secondary flow.

As for the $k$ profile, near the inner curve, the SPH results agree well with the FVM result computed by the standard $k-\epsilon$ model, and both of them over-predict this quantity compared with the experiment.
Near the outer curve, the standard $k-\epsilon$ model obtained a smaller value while the SPH method yielded a slightly bigger one.
This difference is properly due to the low resolution that was used in the FVM simulation.
But in the sub-wall-nearest region, the $k$ value from the SPH method is much closer to the experiment.
And the sudden drop in the wall-nearest region could be ascribed to the fact that the equilibrium assumption of the wall function model is broken due to the secondary flow.
\begin{figure}[htb!]
	\centering
	\includegraphics[trim = 0cm 0cm 11cm 3cm, clip,width=1.0\textwidth]{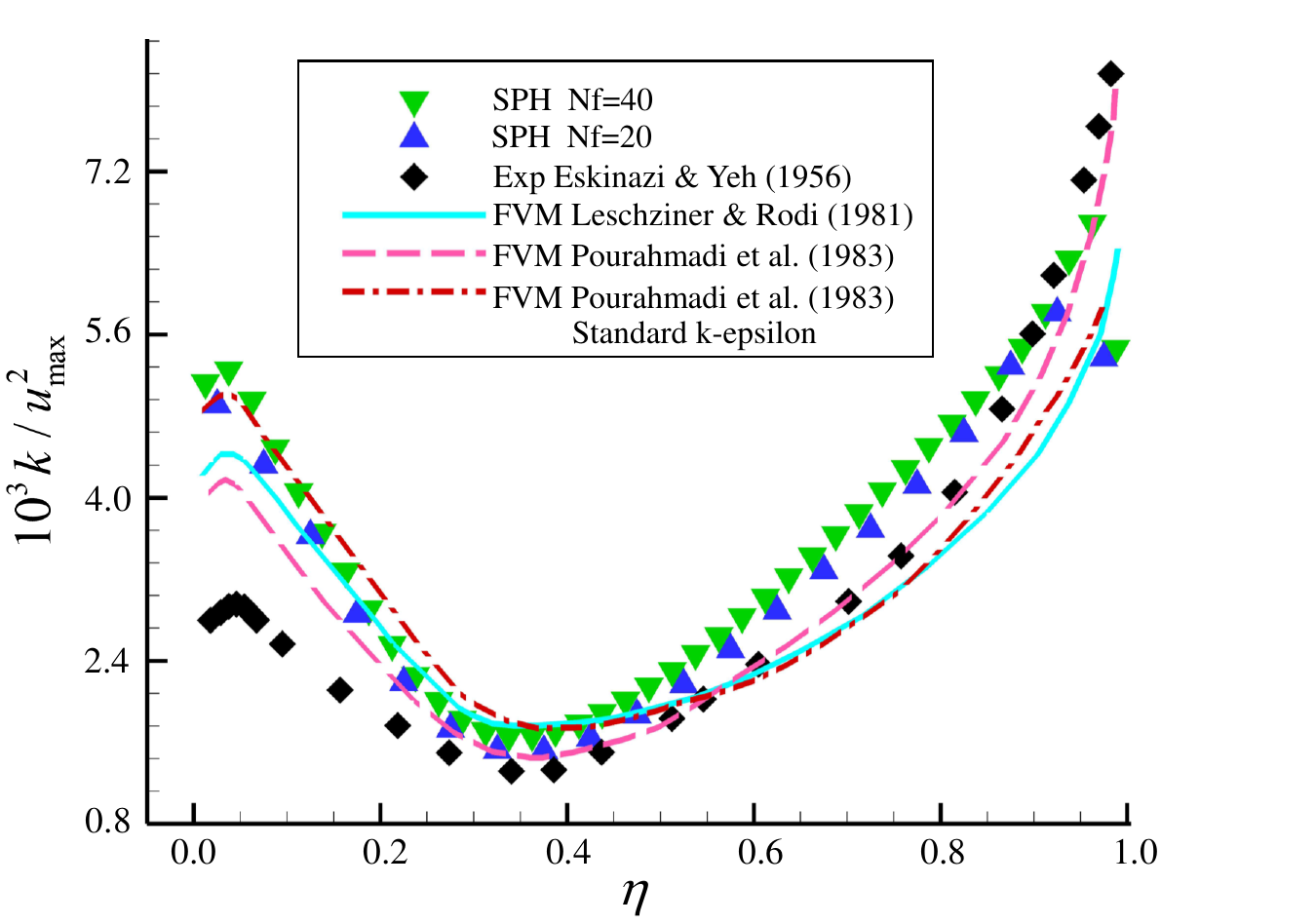}
	\caption{Comparison of the turbulent kinetic energy profiles for the strongly-curved channel.}
	\label{strongly-curved-channel-k-comp}
\end{figure}
%
\subsection{Turbulent Half Converging-Diverging(HCD) channel}
\begin{figure}[htb!]
	\centering
	\includegraphics[trim = 0.0cm 0cm 0cm 11cm, clip,width=1.0\textwidth]{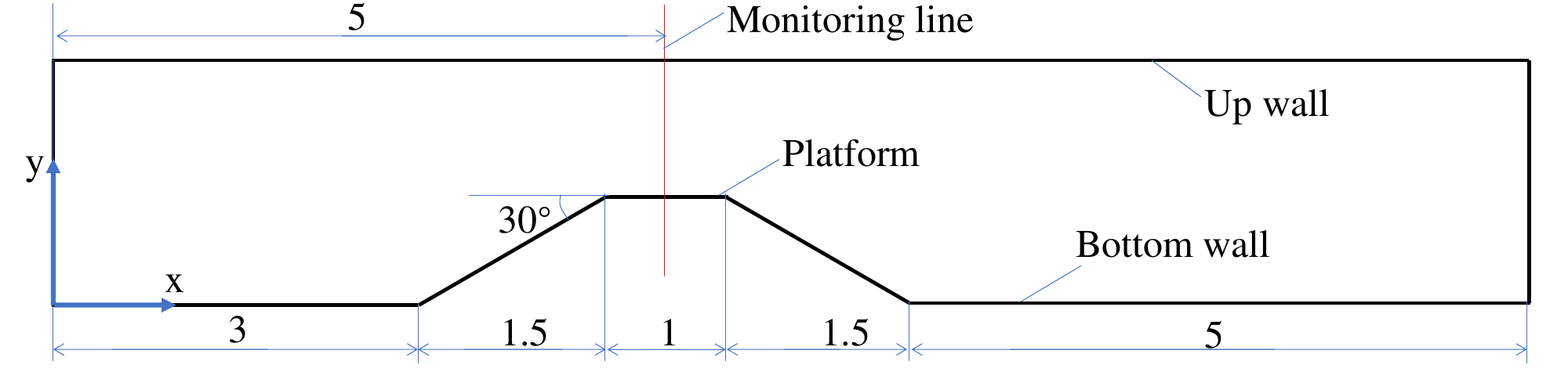}
	\caption{Geometry of the HCD channel and the monitoring position.}
	\label{HCD-GEO}
\end{figure}

To further test the proposed method on the context of gentle flow separation, the turbulent Half Converging-Diverging(HCD) channel is simulated.
The geometry and monitoring line are presented in Fig. \ref{HCD-GEO}.
The channel height is 2 and the resolution is based on this parameter, that means the $N_f$ refers to the number of fluid particles across the channel height.
We compare the cross-sectional velocity on the converging platform because there are complex flow situation while mild flow separation.

\begin{figure}[htb!]
	\centering
	\includegraphics[trim = 0cm 0cm 12cm 4cm, clip,width=0.8\textwidth]{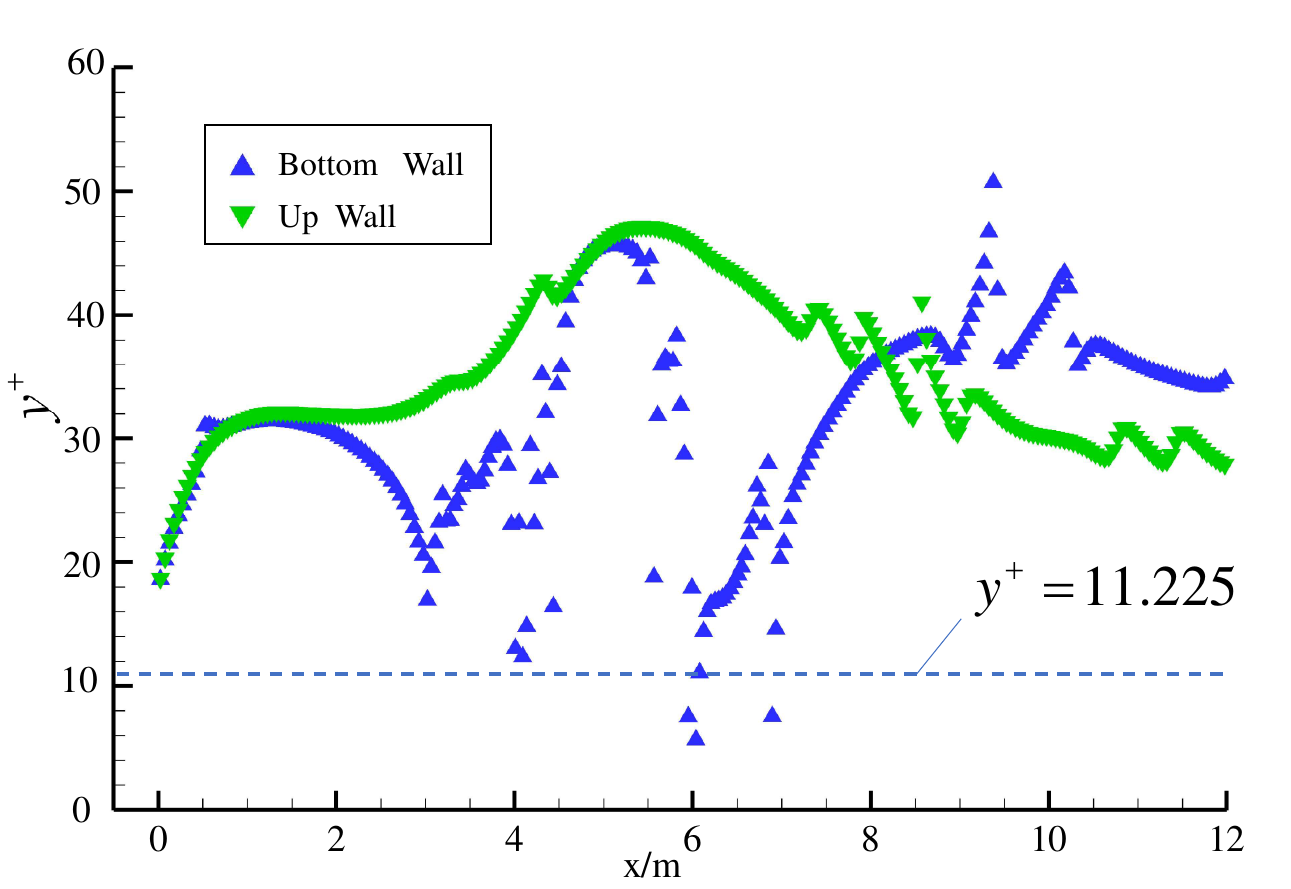}
	\caption{$y^+$ value of the HCD channel from the FVM when $y_p=0.025$.}
	\label{HCD_y_star}
\end{figure}

The Reynolds number is 40000, and the velocity inlet and zero pressure outlet boundary conditions are used.
The simulations are conducted by the FVM and SPH methods.
The mesh-based computations are based on ANSYS FLUENT 16.0 and OpenFOAM-v1912 with the same mesh and conditions.
The distance to wall, $y_p$, is fixed at 0.025.
Figure \ref{HCD_y_star} presents the $y^+$ on the up and bottom wall.
And this value is generally larger than 11.225 (the threshold value between laminar and logarithm laws), meaning that the standard wall function is still effective.

\begin{figure}[htb!]
	\centering
	\includegraphics[trim = 0cm 0cm 0cm 0cm, clip,width=1.0\textwidth]{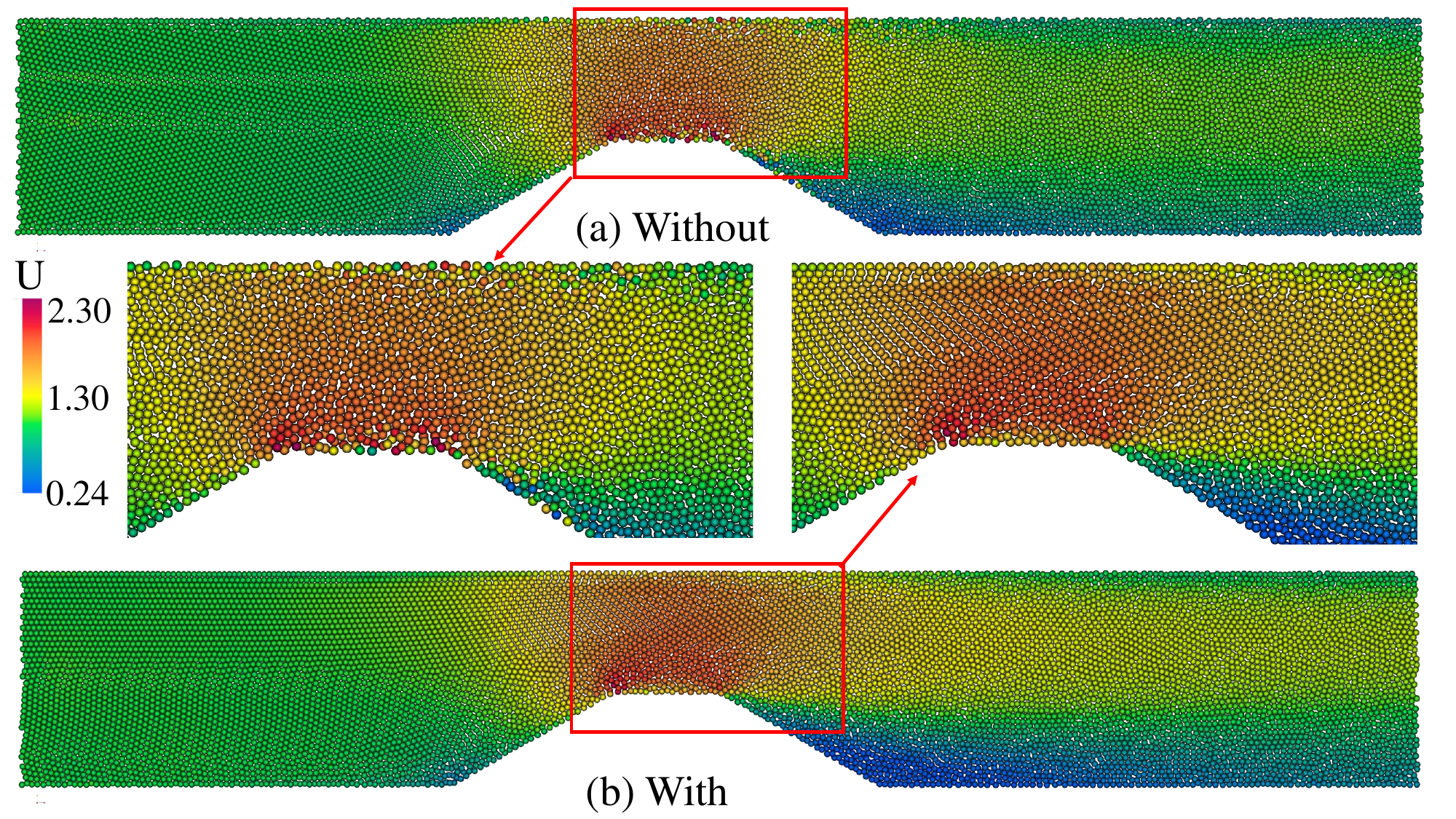}
	\caption{Velocity contours of the HCD channel calculated by the SPH method with or without the ARD technique.}
	\label{HCD-2D-VEL-ARD}
\end{figure}

\begin{figure}[htb!]
	\centering
	\includegraphics[trim = 0.2cm 0cm 0cm 0.1cm, clip,width=1.0\textwidth]{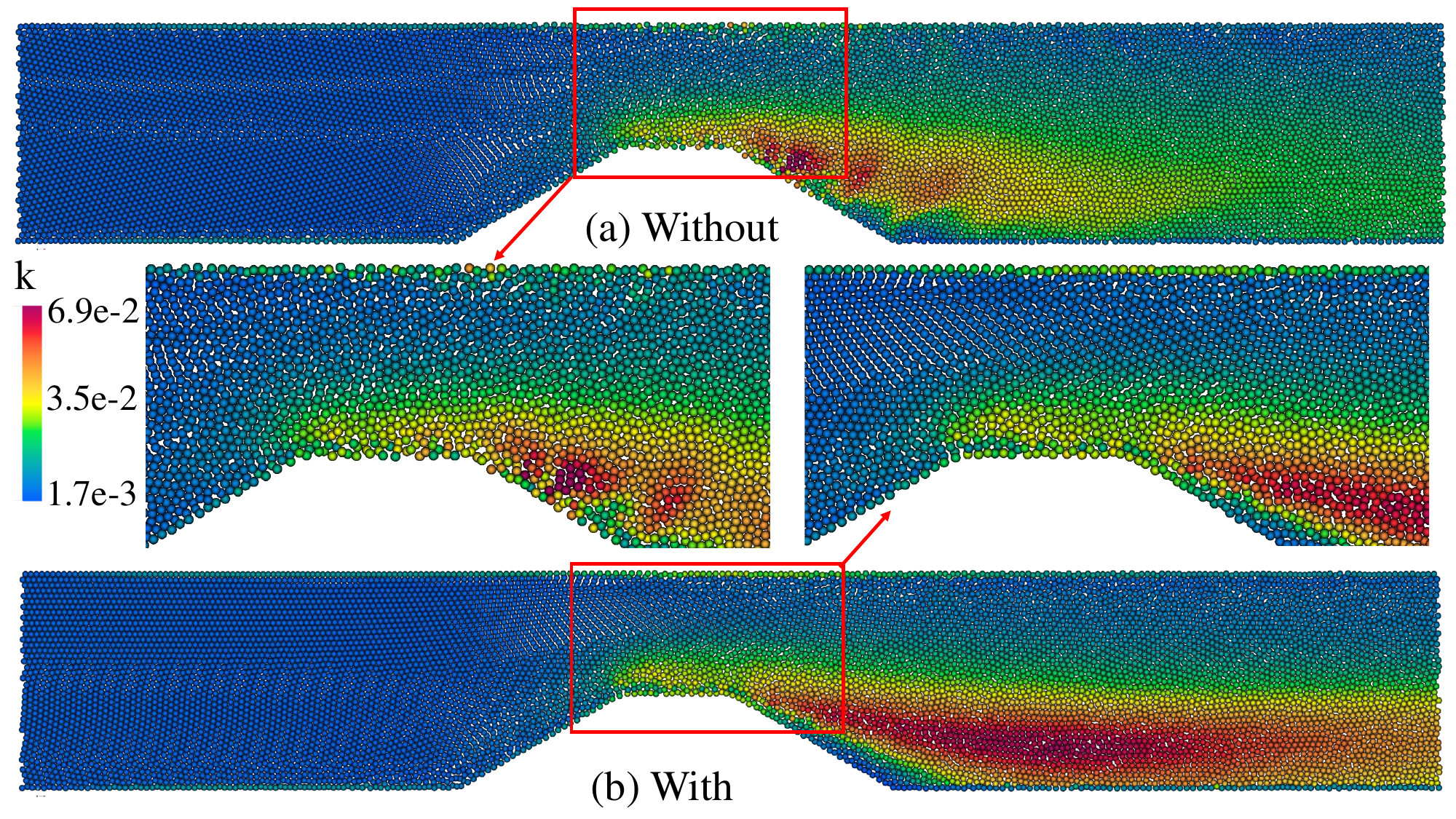}
	\caption{Turbulent kinetic energy contours of the HCD channel calculated by the SPH method with or without the ARD technique.}
	\label{HCD-2D-K-ARD}
\end{figure}

The effectivity of the ARD technique is demonstrated in Fig. \ref{HCD-2D-VEL-ARD} and \ref{HCD-2D-K-ARD}.
Without it, the particles are prone to cluster in the intense shear regions including the front and rear edges of the platform.
The velocity field, shown in Fig. \ref{HCD-2D-VEL-ARD} (a), suffers a strong oscillation, particularly for the particles near the wall, and extremely high and low values can be observed.
The voids appear and wall-adjacent particles become no longer body-fitted, which not only deteriorates the stability but also reduces the accuracy.
The contour of the turbulent kinetic energy exhibits a stronger disturbance, as shown in Fig. \ref{HCD-2D-K-ARD} (a).
Very high values continuously show up behind the platform.
This is due to the "particle vortex" that is explained in the section \ref{ARD}.

Figure \ref{HCD-2D-VEL-ARD} (b) and \ref{HCD-2D-K-ARD} (b) show the results with the ARD scheme.
The velocity field becomes relatively smooth and the local high velocity regions (near the two corners) are well predicted.
A continuous and consistent $k$ profile is displayed.

\begin{figure}[htb!]
	\centering
	\includegraphics[trim = 0cm 0cm 3cm 4cm, clip,width=1.0\textwidth]{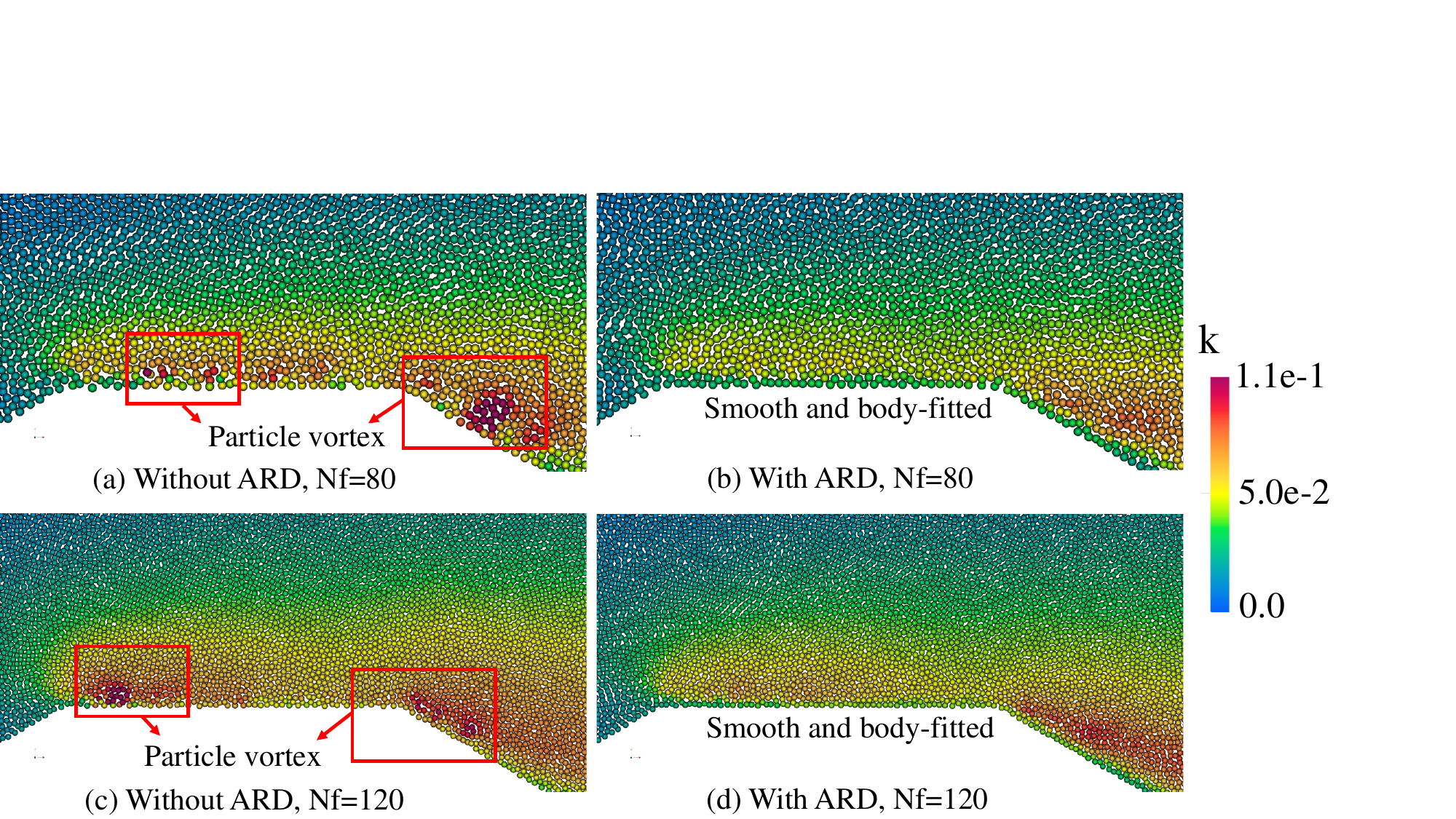}
	\caption{Turbulent kinetic energy in the HCD channel with or without the ARD technique at two additional resolutions}
	\label{HCD-2D-K-ARD-increase-reso}
\end{figure}

Besides, it is worth noted that without the ARD technique , increasing the resolution does not mitigate the disturbance on velocity and $k$, but amplify it, as shown in Fig \ref{HCD-2D-K-ARD-increase-reso}.
Because it stems from the inconsistency between the Lagrangian method and the RANS model.
The RANS model with the wall treatment, although largely reduce the computational effort, results in a discontinuity between the inner fluid domain and the unresolved near wall domain.
For the traditional mesh method, with the cell fixed, it only exhibits the variable jump in the wall model.
However, the Lagrangian method, which naturally has lower numerical dissipation and physical particle movement and shear, exacerbates this discontinuity problem.
The ARD technique is hence proposed to bridge this inconsistency.

To further show the ability of this technique on automatically capturing the strong shear region, an ARD indicator is developed and it is calculated by
\begin{equation}
	N_i^A=\sum_{j} \frac{\max( \rho \beta_{ij} h-\tilde{\mu},0 )}{\rho \beta_{ij} h-\tilde{\mu}}.
	\label{ARD-INDEX}
\end{equation}

\begin{figure}[htb!]
	\centering
	\includegraphics[trim = 0cm 0cm 1cm 1cm, clip,width=1.0\textwidth]{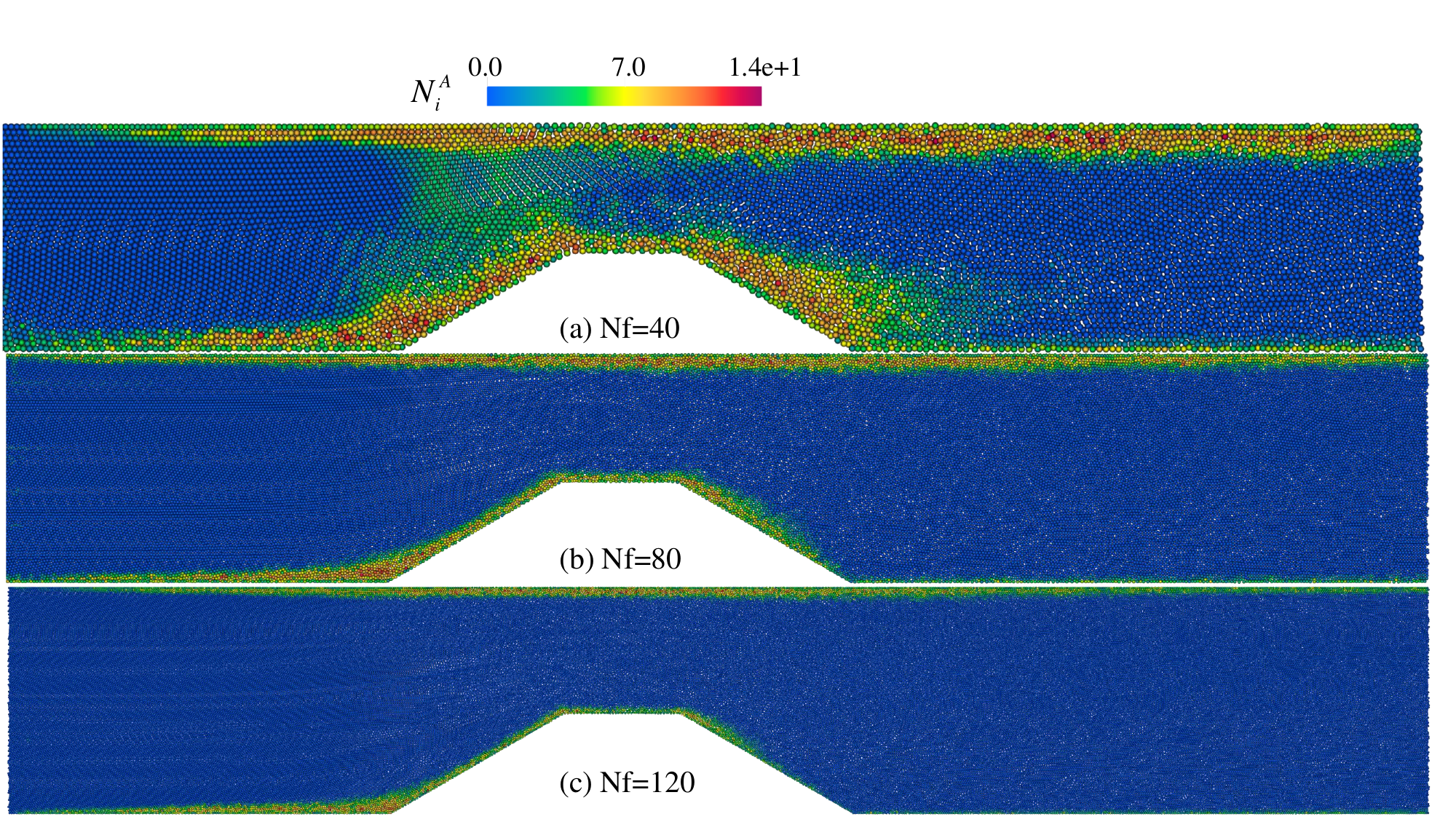}
	\caption{Distribution of the ARD indicator at different resolutions.}
	\label{HCD-ARD-INDICATOR}
\end{figure}

Figure \ref{HCD-ARD-INDICATOR} demonstrates the imposing area and degree of the adaptive dissipation under different resolutions.
The brighter color indicates a stronger degree of the dissipation applied.
And the regions with strong shear, such as the converging segment, back edge of the platform and near wall parts, are accurately captured.
The numerical dissipation gradually decreases with refinement, proving the consistency of this technique.

\begin{figure}[htb!]
	\centering
	\includegraphics[trim = 0cm 0cm 4cm 2cm, clip,width=1.0\textwidth]{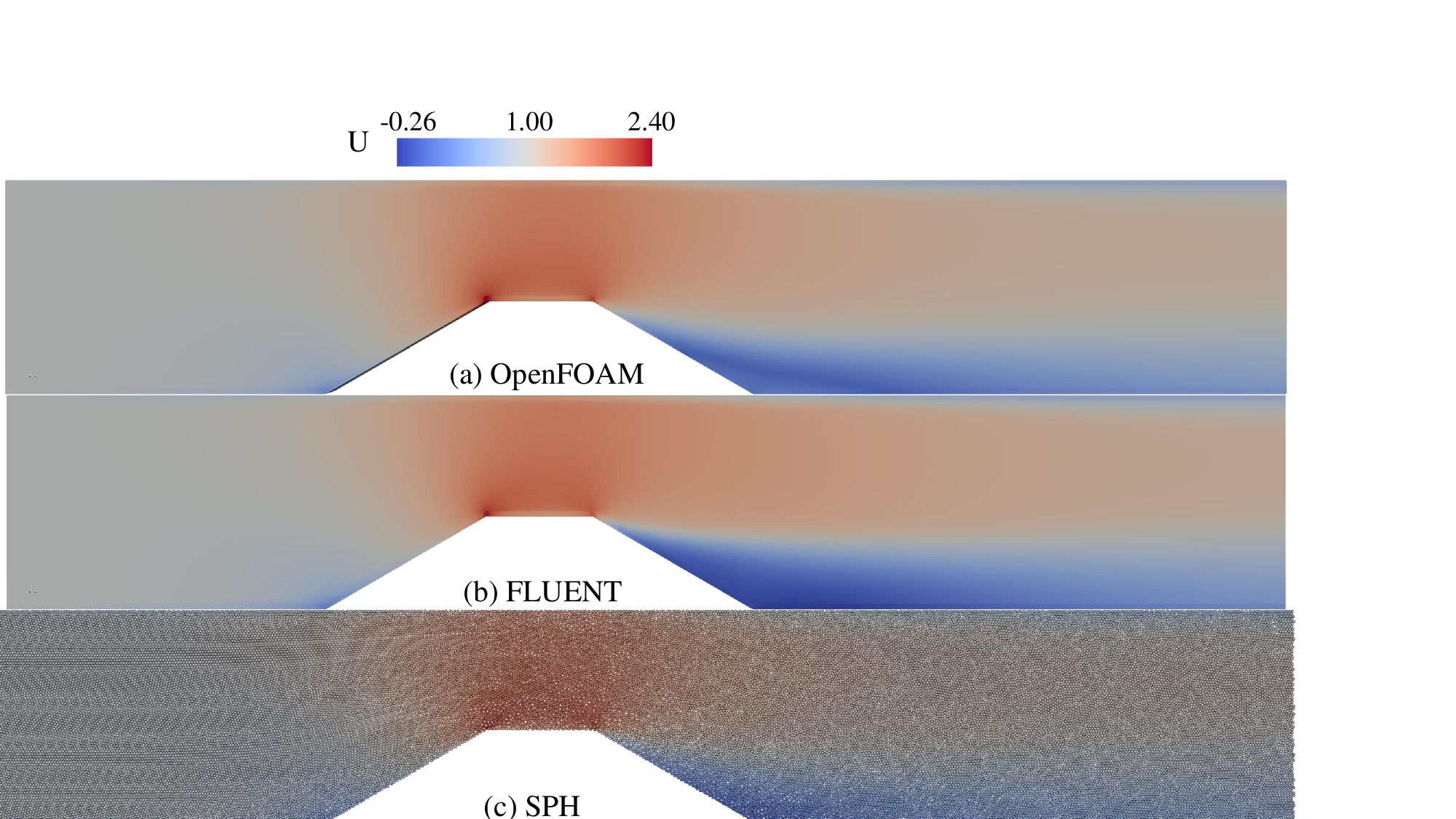}
	\caption{Velocity contour comparison between the FVM and SPH methods.}
	\label{HCD-VEL-NF80}
\end{figure}

\begin{figure}[htb!]
	\centering
	\includegraphics[trim = 0.0cm 0cm 14cm 4cm, clip,width=0.9\textwidth]{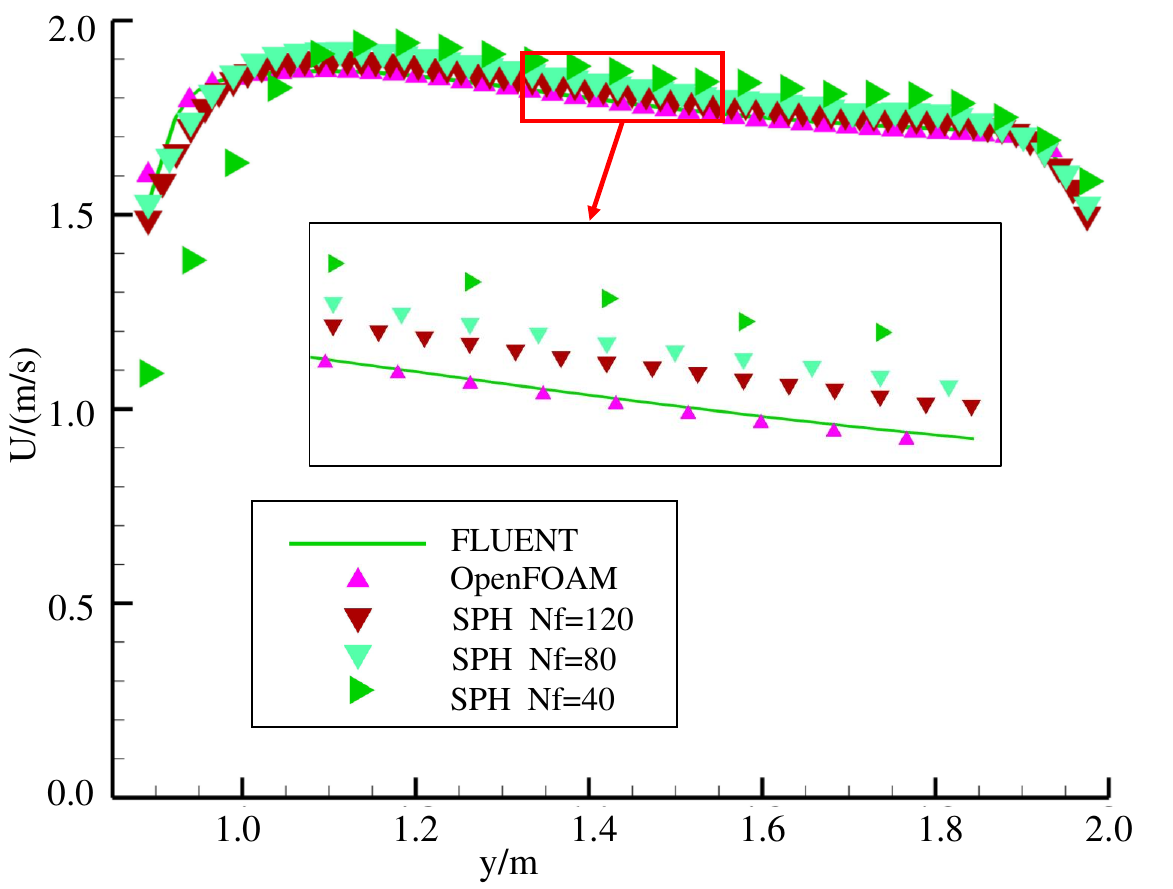}
	\caption{Velocity comparison between the FVM and SPH methods on the monitoring line.}
	\label{HCD-VEL-line}
\end{figure}

\begin{figure}[htb!]
	\centering
	\includegraphics[trim =0.0cm 0cm 15cm 5cm, clip,width=0.9\textwidth]{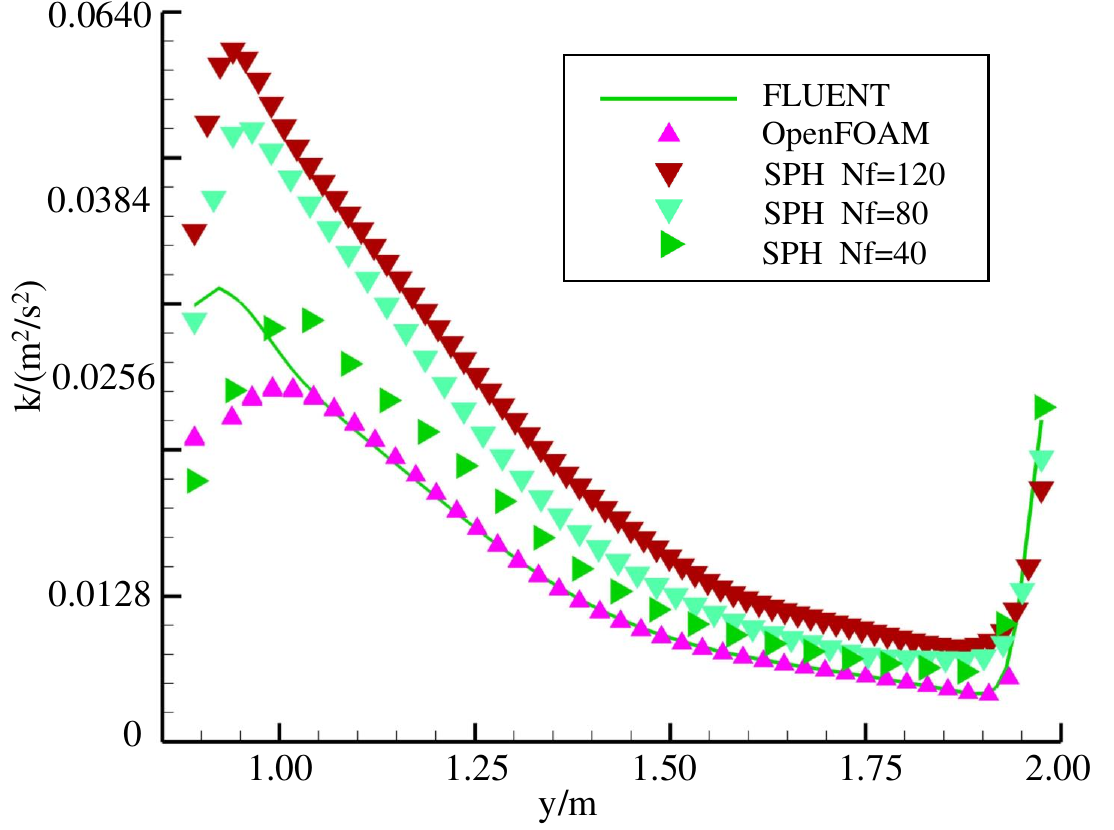}
	\caption{Turbulent kinetic energy comparison between the FVM and SPH methods on the monitoring line.}
	\label{HCD-K-line}
\end{figure}

Finally, we compare the results calculated by proposed method and FVM.
The velocity contour comparison at a moderate resolution ($N_f=80$) is shown in Fig. \ref{HCD-VEL-NF80}.
Both the two methods predict very smooth field, and the negative velocity behind the platform is clearly presented.
The quantitative comparisons of the velocity and $k$ are shown in Fig \ref{HCD-VEL-line} and \ref{HCD-K-line}.
With the offset model activated and the $y_p$ fixed, we also conduct the convergence test for this case with 3 different resolutions.
Both velocity and turbulent kinetic energy computed by the SPH method show very good convergence, and they agree well with the results from the FVM.
Besides, since this case is highly nonlinear, a small disturbance could make the simulation converge to another solution.
Therefore, occasionally, with different initial particle relaxation distribution, a different vortex shape would appear.
But in most cases, the mentioned results in this article are promising.



\section{Conclusion}
\label{section-conclusion}
In this paper, we propose a WCSPH method for solving the wall-bounded turbulent flow 
without or with gentle flow separation.
The k-$\epsilon$ RANS equations with the wall model are discretized within the state-of-the-art Riemann WCSPH framework to ensure numerical stability and accuracy. 
To address the Lagrangian challenges of the RANS model, 
such as intense particle shear and near-wall discontinuities, 
we introduce specialized techniques for both the main stream and near-wall regions. 
The main stream improvements include the adaptive Riemann-eddy dissipation and the limited transport velocity formulation.
While the former solves the over-damping issue by switching the Riemann dissipation 
and eddy viscosity adaptively,
the latter addresses the over-prediction of the turbulent kinetic energy $k$, 
particularly at high resolution, by suppressing the transport velocity correction when consistency residue is negligible.
Furthermore, we find the $k$ over-estimation problem in particle-based methods is mainly attributed to two aspects, the particle vortex caused by the inconsistency and the over-correction in the turbulent plug flow region.

For the near-wall treatments, 
we propose a Lagrangian-meshless implementation of the wall model, in which the wall dummy particles are still used. 
The key improvements are: 
(1) developing a general wall-model-consistent calculation scheme for the wall shear stress and velocity gradients for complex geometries using tangential flow velocity and direction, 
(2) introducing four wall boundary conditions to handle governing 
and transport equation terms, and address the truncation and sub-truncation problems, 
(3) proposing a constant-$y_p$ strategy to increase the stability,
and boundary-offset technique to achieve the rigorous convergence test. 

The method is verified through four wall-bounded turbulent flow cases 
in straight, curved, and converging-diverging channels, 
demonstrating excellent agreement with analytical solutions, 
mesh-based methods, and experimental data. 
Future work will extend this WCSPH-RANS method 
to additional RANS models and turbulent fluid-structure interaction simulations.

\addcontentsline{toc}{section}{Acknowledgement}

%
%

\bibliographystyle{elsarticle-num}
\bibliography{FirstPaper}
%
%
\end{document}